\documentclass[AMS,STIX1COL,SYSTEMFONTS]{WileyNJD-v2}

\usepackage{multicol} % Used for the two-column layout of the document
\usepackage{float} % Required for tables and figures in the multi-column environment - they need to be placed in specific locations with the [H] (e.g. \begin{table}[H])
\usepackage{hyperref} % For hyperlinks in the PDF
\usepackage{paralist} % Used for the compactitem environment which makes bullet points with less space between them
\usepackage{float}
\usepackage{lipsum} % Package to generate dummy text throughout this template
\usepackage{blindtext}
\usepackage{graphicx} 
\usepackage{caption}
\usepackage{natbib, url}
\usepackage{subfig}
\usepackage{soul}
\usepackage{anyfontsize}

\articletype{Article Type}%

\received{XXXXX}
\revised{XXXXX}
\accepted{XXXX}

\newcommand{\given}{\,|\,}

\graphicspath{{./piceps/}}

\begin{document}

\title{Practical Bayesian Modeling and Inference for Massive Spatial Datasets On Modest Computing Environments\protect\thanks{Massive Spatial Data Analysis On Modest Computing Environments}}

\author[1]{Lu Zhang}

\author[2]{Abhirup Datta}

\author[3]{Sudipto Banerjee*}

\authormark{Lu Zhang \textsc{et al}}

\address[1]{\orgdiv{Department of Biostatistics}, \orgname{University of California Los Angeles}, \orgaddress{\state{California}, \country{USA}}}

\address[2]{\orgdiv{Department of Biostatistics}, \orgname{Johns Hopkins University}, \orgaddress{\state{Maryland}, \country{USA}}}

\address[3]{\orgdiv{Department of Biostatistics}, \orgname{University of California Los Angeles}, \orgaddress{\state{California}, \country{USA}}}

\corres{*Sudipto Banerjee, UCLA Department of Biostatistics. \email{sudipto@ucla.edu}}

\presentaddress{650 Charles E. Young Drive South, Los Angeles, CA 90095-1772.}

\abstract[Summary]{With continued advances in Geographic Information Systems and related computational technologies, statisticians are often required to analyze very large spatial datasets. This has generated substantial interest over the last decade, already too vast to be summarized here, in scalable methodologies for analyzing large spatial datasets. Scalable spatial process models have been found especially attractive due to their richness and flexibility and, particularly so in the Bayesian paradigm, due to their presence in hierarchical model settings. However, the vast majority of research articles present in this domain have been geared toward innovative theory or more complex model development. Very limited attention has been accorded to approaches for easily implementable scalable hierarchical models for the practicing scientist or spatial analyst. This article devises massively scalable Bayesian approaches that can rapidly deliver inference on spatial process that are practically indistinguishable from inference obtained using more expensive alternatives. A key emphasis is on implementation within very standard (modest) computing environments (e.g., a standard desktop or laptop) using easily available statistical software packages. %\textcolor{orange}{\sout{without requiring message-parsing interfaces or parallel programming paradigms.}} 
Key insights are offered regarding assumptions and approximations concerning practical efficiency.}


% \jnlcitation{\cname{%
% \author{Zhang L.}, 
% %\author{A. Datta}, 
% %\author{S. Banerjee}, 
% \author{A. Datta}, and 
% \author{S. Banerjee}} (\cyear{2018}), 
% \ctitle{Practical Bayesian Modeling and Inference for Massive Spatial Datasets On Modest Computing Environments \protect\thanks{Massive Spatial Data Analysis On Modest Computing Environments}}, \cjournal{Statistical Learning and Data Mining}, \cvol{2018; 0:1--15}.
% }

\maketitle

%\footnotetext{\textbf{Abbreviations:} ANA, anti-nuclear antibodies; APC, antigen-presenting cells; IRF, interferon regulatory factor}

\section{Introduction}\label{sec: intro}

Rapidly increasing usage and growing capabilities of Geographic Information Systems (GIS) have spawned considerable research in modeling and analyzing spatial datasets in diverse disciplines including, but not limited to, environmental sciences, economics, biometry and so on \citep[see, e.g.,][]{gelfand2010handbook, cressie2015statistics, banerjee2014hierarchical}. Much of spatial modeling is carried out within the familiar hierarchical modeling paradigm,
\begin{equation}\label{eq: generic_paradigm}
[\mbox{data}\given \mbox{process}] \times [\mbox{process}\given \mbox{parameters}] \times [\mbox{parameters}]\; .
\end{equation}
For point-referenced data sets, where spatial locations are indexed by coordinates on a map, the ``process'' is modeled as a spatial random field over the domain of interest and the observations are treated as a finite realization of this random field. The Gaussian process (GP) is, perhaps, the most conspicuous of process specifications and offers flexibility and richness in modeling. 
%It is denoted as $\{w(s) \sim GP(m_\theta(\cdot), C_\theta(\cdot, \cdot)), s \in D\}$, where $D$ is the domain, $\theta$ is a set of unknown parameters, $m_\theta(\cdot)$ is a mean function defining the trend, and $C_\theta(\cdot, \cdot)$ is a positive definite covariance function. 
The GP's popularity as a modeling tool is enhanced due to their extensibility to multivariate and spatial-temporal geostatistical settings, although we do not pursue such generalizations in this article. They also provide comparatively greater theoretical tractability among spatial processes \citep{stein99}.  

Fitting GPs incur onerous computational costs that severely hinders their implementation for large datasets. The key bottleneck stems from the massive spatial covariance matrix present in the multivariate normal density for the finite realizations of the GP. For irregularly situated spatial locations, as is common in geostatistics, these matrices are typically dense and carry no exploitable structure to facilitate computations. 
%To be precise, if $w(S)$ is the $n\times 1$ vector of $w(s_i)$'s over a set $S = \{s_1, s_2, \ldots, s_n\}$ of $n$ locations. Then $w(S) \sim N(m_\theta(S), C_{\theta}(S,S))$, where $m_\theta(S)$ is the corresponding $n\times 1$ mean vector and $C_{\theta}(S,S)$ is an $n\times n$ covariance matrix whose entries are given by the covariance function $C_{\theta}(s_i,s_j)$. Computing this density requires the Cholesky decomposition for $C_{\theta}(S,S)$ involving storage and floating point operations in the order of $\sim n^3$, usually performed in each iteration of the fitting algorithm. 
Even for a modestly large number of points ($\approx 50,000$ or greater), the computational demands become prohibitive for a modern computer and preclude inference from GP models.  

A substantial literature exists on methodologies for massive spatial datasets and it is already too vast to be summarized here \citep[see, e.g.,][and references therein]{banerjee2017high, heatoncontest2017}. Some are more amenable than others to the hierarchical setup in (\ref{eq: generic_paradigm}). Even within the hierarchical paradigm, there is already a burgeoning literature on massively scalable spatial process models. There are two pressing issues facing the practicing spatial analyst. The first is to analyze massive amounts of spatial data on ``modest'' computing environments such as standard desktop or laptop architectures. %\textcolor{orange}{\sout{By ``modest computing environments'', we mean  environments such as \texttt{R} (\url{http://cran.r-project.org}) or \texttt{STAN} (\url{http://mc-stan.org}) that do not require knowledge of lower-level languages or distributed programming paradigms.}} 
The second pressing issue is that of \emph{full inference} that subsumes parameter estimation, spatial prediction of the outcome, and estimation of the underlying latent process. Yet the size of the datasets easily exceed the CPU memory available for computing, which means that we need to rely upon statistical models that will enable analysis with the available memory.   

Some scalable processes such as the multi-resolution predictive process models proposed by \cite{katzfussmultires} or the nearest-neighbor Gaussian process (NNGP) models by \cite{datta16} can be programmed in modest computing environments to estimate parameters and predict outcomes, but not necessarily infer on the latent process efficiently. \cite{katzfussmultires} does not address this, while \cite{datta16} and \cite{datta16b} implement high-dimensional Gibbs sampling algorithms that had to be run for several iterations on a high-performance computing environment to yield adequate convergence due to high autocorrelations. Other approaches such as Gaussian Markov random field (GMRF) approximations to spatial processes \cite{ruemartinochopin2009, lindgrenruelindstrom2011} use Integrated Nested Laplace Approximations (INLA) for computing the marginal distribution of the process at given locations. These approximations can be implemented on standard environments for a variety of spatial models using the \texttt{R-INLA} software (\url{www.r-inla.org}). This is computationally more promising than MCMC, but is still an iterative procedure requiring convergence assessment. Its performance is yet to be demonstrated for analyzing massive spatial data with millions of spatial locations on modest computing environments.       

This article outlines strategies for achieving fully model-based Bayesian inference including parameter estimation, response surface predictions and interpolation of the latent spatial process for massive spatial datasets on modest computing environments. %This manuscript is designed for the \emph{Practice} section of this journal. The focus is not on novel statistical methodologies, nor is it on proposing new highly complex models. Instead, we seek to offer some effective strategies for the practicing spatial analyst to run robust and quick analysis of massive spatial datasets on very standard desktop or laptop architectures. 
To achieve this goal, we need a massively scalable spatial process that will be able to estimate (\ref{eq: generic_paradigm}) by obviating the memory obstacles. Here, there are a few choices that are well-suited for (\ref{eq: generic_paradigm}) all of whom seem to be competitive based upon the recent ``contest'' paper by \cite{heatoncontest2017}, but we opt for the sparsity-inducing Nearest-neighbor Gaussian process (NNGP) primarily because of its ease of use and also because of its easier accessibility through the \texttt{spNNGP} package available from \texttt{cran.r-project.org/web/packages/spNNGP} (see Section~\ref{sec: nngp_review}). 

%\sout{In fact, \cite{finley2017applying} outline several strategies for estimating NNGP models, including a conjugate Bayesian NNGP response model for exact inference without requiring MCMC that was fitted to a dataset with approximately 5 million locations in a matter of seconds. However, the approach did not accommodate the latent process and, hence, did not generalize to (\ref{eq: generic_paradigm}). For the latent process models of \cite{datta16}, \cite{finley2017applying} used the collapsed model after integrating out the latent effects and exploited permutation-based sparse Cholesky methods embedded within MCMC algorithms. This still proved too expensive for massive datasets in the order of $10^6$ locations. Our current contribution lies in casting the latent process models of \cite{datta16} within a conjugate Bayesian framework for exact inference avoiding MCMC. Like \cite{finley2017applying} we will exploit conjugacy in conjunction with cross-validatory estimation of process parameters, but we need to be careful that the model formulation and computations will not require loading large data objects into memory at any point. In addition to the latent process, we also show how the response model, also analyzed by \cite{finley2017applying} on much more sophisticated computing environments, can be executed on very modest architectures. The details of the conjugate Bayesian formulation and algorithms for their effective implementation constitute the novelty of this paper.}    
%\textcolor{blue}
{In fact, \citet{finley2017applying} outlines several strategies for estimating NNGP models, including a conjugate response NNGP model and a collapsed NNGP model.  The conjugate response NNGP model can provide exact inference without requiring MCMC and has been demonstrated to effectively fit a dataset with approximately 5 million locations in a matter of seconds on a Linux workstation. % with two 18-core Intel processors and 512 GB of memory.
However, the response model does not accommodate the latent process and, hence, is restrictive in its inferential capabilities compared to (\ref{eq: generic_paradigm}). The collapsed NNGP model, on the other hand, is embedded within MCMC algorithms and is able to provide the posterior inference of the latent process. It can exploit permutation-based sparse Cholesky methods, but the approach requires specialized libraries and can still be too expensive for massive datasets in the order of $10^6$ locations for standard computing environments. We briefly introduce the conjugate response NNGP model in section~\ref{se: conjugate_nngp_response}, and the discussion of the collapsed NNGP model can be found in section~\ref{sec: nngp_latent}. Our contribution lies in casting the latent process models of \cite{datta16} within a conjugate Bayesian framework for exact inference so as to avoid MCMC while being able to achieve full Bayesian inference including estimation of the latent process. We propose a conjugate latent NNGP model that exploits conjugacy in conjunction with cross-validatory estimation of a small set of process parameters, and the model formulation and computations will not require loading large data objects into memory at any point, allowing fitting for massive datasets in the order of $10^6$ on computer environments like standard desktop or laptop architecture. The details of this Bayesian formulation and the algorithms for their effective implementation constitute the novelty of this paper.}

The remainder of the paper evolves as follows. Section~\ref{sec: nngp_review} provides a brief review of nearest-neighbor Gaussian process and NNGP based models. Section~\ref{sec: conj_Bayesian_model} develops the conjugate NNGP based models, emphasizing the conjugate latent NNGP model, and devises algorithms for practical implementation. A simulation study is presented in Section~\ref{sec: simulation_study} for discussing the performance of the proposed models, while an analysis on sea surface temperature with over 2.5 million locations is conducted in Section~\ref{sec: real_data_analysis}. %All computations for the proposed framework have been carried out in \texttt{R}.
Finally, we conclude with some discussion in Section~\ref{sec: conclusion}.

\section{The nearest-neighbor Gaussian process}\label{sec: nngp_review}

\noindent The computational burden in GP models arises from the $n\times n$ covariance matrix $C_{\theta}(S,S)$, where $S = \{s_1,s_2,\ldots,s_n\}$ is the set of observed locations. %\textcolor{orange}{\sout{and $n$ is large.}} 
The $(i,j)$-th element of this matrix is the value of a spatial covariance function evaluated at locations $s_i$ and $s_j$. Spatial covariance functions in general do not produce exploitable structures in the resulting matrices. One effective approach to achieve efficient computations is to replace $C_{\theta}(S,S)$ with an approximate $\tilde{C}_{\theta}(S,S)$ such that the inverse of $\tilde{C}_{\theta}(S,S)$ is sparse. There are multiple options, but notable among them are approximations based upon Gaussian Markov random fields or GMRFs \citep[see, e.g.,][]{rueheld04,ruemartinochopin2009} that yield computationally efficient sparse representations. An alternative approach exploits an idea familiar in graphical models or Bayesian networks \citep[see, e.g.,][]{lauritzen96,bishop2006,murphy2012} that has also been exploited by \cite{ve88}, \cite{stein04} and \cite{stroud17} to construct composite likelihoods for inference. 
%Recently, \citet{stroud17} has exploited this idea to propose preconditioned conjugate gradient algorithms for Bayesian and maximum likelihood estimates on large incomplete lattices. 
\citet{datta16,datta16b} extended this idea to construct a Nearest Neighbor Gaussian Process (NNGP) for modeling large spatial data. NNGP is a well defined Gaussian Process that yields finite dimensional Gaussian densities with sparse precision matrices. It delivers massive scalability both in terms of parameter estimation and spatial prediction or ``kriging''.

%In terms of inferential performance and computational efficiency, there do not appear to be substantial differences between the GMRF approach and the Vecchia-based approaches such as the NNGP. While each approach probably has some advantages and disadvantages compared to the other, in terms of practical inference there is little to choose between them so we do not embark upon such comparisons here. However, while highly specialized algorithms such as the Integrated Nested Laplace Approximation (INLA) developed by \citet{ruemartinochopin2009} can be effectively used for GMRF-based approximate models, there is little guidance currently available for implementing full and exact Bayesian inference for NNGP models. Here, we focus upon them.  

\subsection{Response NNGP model}\label{sec: nngp_response}

Consider modeling a point-referenced outcome as a partial realization of a Gaussian process, $\{y(s) : s\in D\} \sim GP(m_{\theta}(s), C_\theta(\cdot, \cdot))$ on a spatial domain $D \in \Re^d$. The mean and covariance functions are assumed to be determined by one or more parameters in a set $\theta$. The finite-dimensional distribution for the $n\times 1$ vector $y(S)$ with elements $y(s_i)$ is multivariate normal with mean $m_{\theta}(S)$ and covariance matrix $C_{\theta}(S,S)$. 
As a directed acyclic graph (DAG) \citep[][]{bishop2006}, the joint density is $\displaystyle p(y(S)) = \prod_{i=1}^n p(y(s_i) \given y(\mbox{Pa}[s_i])$,
% \[
% p(y(S)) = p(y(s_1))\prod_{i=2}^n p(y(s_i) \given y(s_1),\ldots, y(s_{i-1})) = \prod_{i=1}^n p(y(s_i) \given y(\mbox{Pa}[s_i])\;,
% \]
where $\mbox{Pa}[s_1]$ is the empty set and $\mbox{Pa}[s_i] = \{s_1,s_2,\ldots,s_{i-1}\}$ for $i=2,3,\ldots,n-1$ is the set of parent nodes with directed edges to $s_i$. \cite{ve88} suggested approximating the multivariate normal likelihood by shrinking $\mbox{Pa}[s_i]$ from the set of all nodes preceding $s_i$ to a much smaller subset of locations preceding $s_i$ that are among the $m$ (a fixed small number) nearest neighbors of $s_i$ based upon their Euclidean distance. \cite{datta16} extended that notion to arbitrary points in the domain by defining
% \[
% \resizebox{1\hsize}{!}{$
% \mbox{Pa}[s] =
% \left\{\begin{array}{ll}
% \mbox{empty set} & \mbox{ if $s=s_1$}\;,\\
% \{s_1, s_2, \ldots, s_{i-1}\} & \mbox{ if $s\in S$ and $i \leq m$}\;, \\
% \mbox{$m$ closest points to $s$ among $\{s_j\}_{j < i}$} & \mbox{ if $s\in S$ and $i > m$}\;, \\
% \mbox{$m$ closest points to $s$ among $S$} & \mbox{ if $s \notin S$}\;.
% \end{array}
% \right. $} 
% \]
\[
\mbox{Pa}[s] = \left\{\begin{array}{ll}
\mbox{empty set} & \mbox{ if $s=s_1$}\;,\\
\{s_1, s_2, \ldots, s_{i-1}\} & \mbox{ if $s\in S$ and $i=1,2,\ldots,m$}\;, \\
\mbox{$m$ closest points to $s$ among $\{s_1, s_2, \ldots, s_{i-1}\}$} & \mbox{ if $s\in S$ and $i > m$}\;, \\
\mbox{$m$ closest points to $s$ among $S$} & \mbox{ if $s \notin S$}\;.
\end{array}
\right.  
\]
for any arbitrary point $s$ in the domain, where $m$ is the fixed number of nearest neighbors. This results in another multivariate Gaussian density   
% \begin{equation}\label{eq: nngp_likelihood}
% \begin{aligned}
% p(y(S)) &= N(y(S)\given m_{\theta}(S), C_{\theta}(S,S)) \\
% &\approx N(y(S)\given m_{\theta}(S), \tilde{C}_{\theta}(S,S))\; , 
% \end{aligned}
% \end{equation}
\begin{equation}\label{eq: nngp_likelihood}
\begin{aligned}
p(y(S)) &= N(y(S)\given m_{\theta}(S), C_{\theta}(S,S)) \approx N(y(S)\given m_{\theta}(S), \tilde{C}_{\theta}(S,S))\; , 
\end{aligned}
\end{equation}
where $\tilde{C}_{\theta}(S,S)^{-1} = (I-A_S)^{\top}D_S^{-1}(I-A_S)$ is sparse, $A_S$ is sparse and strictly lower triangular with $A_S(i,i)=0$ for $i=1,2,\ldots,n$ and at most $m$ non-zero entries in each row, and $D_S$ is diagonal whose elements are the conditional variances  $\mbox{var}\{y(s_i)\given y(\mbox{Pa}[s_i])\}$ based upon the full GP model, i.e., $D_S(1,1) = C_{\theta}(s_1,s_1)$ and 
$D_S(i,i) = C_{\theta}(s_i,s_i) - C_{\theta}(s_i,\mbox{Pa}[s_i])C_{\theta}(\mbox{Pa}[s_i],\mbox{Pa}[s_i])^{-1}C_{\theta}(\mbox{Pa}[s_i],s_i)$ for $i = 2, \ldots, n$.
Turning to the structure of $A_S$, all its elements are completely determined from $C_{\theta}(S,S)$. Its first row, i.e., $A_S(1,)$ has all zeroes. For the $i+1$-th row, the nonzero entries appear in the positions indexed by $\mbox{Pa}[s_{i+1}]$ and are obtained as row vectors, 
\[
A_S(i+1,\mbox{Pa}[s_{i+1}]) = C_{\theta}(s_{i+1},\mbox{Pa}[s_{i+1}])C_{\theta}(\mbox{Pa}[s_{i+1}],\mbox{Pa}[s_{i+1}])^{-1}\;.
\]
%These equations are derived by equating the conditional expectations $\mbox{E}[y(s_{i+1})\given w(\mbox{Pa}[s_{i+1}])]$ in terms of the entries of $A(S)$ and $m_{\theta}(S)$ with that from the multivariate normal density corresponding to the full GP. 
The nonzero entries in each row of $A_S$ are precisely the ``kriging'' weights of $y(s_i)$ based upon the values of $y(s)$ at neighboring locations, i.e., $\mbox{Pa}[s_i]$ \citep{Chiles1999}. The $\tilde{C}_{\theta}(S,S)$, constructed as above, is called an NNGP approximation to $C_{\theta}(S,S)$.  

With the above definition of $\mbox{Pa}[s]$, we can express the partial realizations of an NNGP as a linear model. Let $S$ be the set of the $n$ observed locations as defined earlier (and $n$ is assumed to be large) and let $U = \{u_1, u_2, \ldots, u_{n'}\}$ be a set of $n'$ arbitrary locations where we wish to predict $y(s)$. Then, 
% \begin{equation}\label{eq: nngp_linear_model}
% \resizebox{.9\hsize}{!}{$\underbrace{ \left[\begin{array}{c} y(S) \\ y(U)\end{array}\right]}_{y} = \underbrace{ \left[\begin{array}{c} m_{\theta}(S) \\ m_{\theta}(U) \end{array} \right]}_{m_{\theta}} + \underbrace{\left[\begin{array}{c} A(S) \\ A(U) \end{array}\right]}_{A} (y(S) - m_{\theta}(S)) + \underbrace{\left[\begin{array}{c} \eta(S) \\ \eta(U) \end{array}\right]}_{\eta}\; ,$}
% \end{equation}
\begin{equation}\label{eq: nngp_linear_model}
%\resizebox{.9\hsize}{!}
\underbrace{ \left[\begin{array}{c} y(S) \\ y(U)\end{array}\right]}_{y} = \underbrace{ \left[\begin{array}{c} m_{\theta}(S) \\ m_{\theta}(U) \end{array} \right]}_{m_{\theta}} + \underbrace{\left[\begin{array}{c} A(S) \\ A(U) \end{array}\right]}_{A} (y(S) - m_{\theta}(S)) + \underbrace{\left[\begin{array}{c} \eta(S) \\ \eta(U) \end{array}\right]}_{\eta}\; ,
\end{equation}
where $\displaystyle \eta \sim N\left( \left[ \begin{array}{c} 0\\ 0\end{array}\right], \left[\begin{array}{cc} D(S) & O \\ O & D(U)\end{array} \right] \right)$, $D(U)$ is $n'\times n'$ diagonal and $A(U)$ is sparse $n'\times n$ formed by extending the definitions of $D(S)$ and $A(S)$ as
%\[
% \begin{equation}\label{eq: DU_AU}
% \begin{aligned}
% D(U)&(i,i) = C_{\theta}(u_i,u_i) - \\   &C_{\theta}(u_i,\mbox{Pa}[u_i])C_{\theta}(\mbox{Pa}[u_i],\mbox{Pa}[u_i])^{-1}C_{\theta}(\mbox{Pa}[u_i],u_i)\;,\\
% % \]
% %\[
% A(U)&(i,\mbox{Pa}[u_{i}]) = C_{\theta}(u_{i},\mbox{Pa}[u_{i}])C_{\theta}(\mbox{Pa}[u_{i}],\mbox{Pa}[u_{i}])^{-1}\;.
% %\]
% \end{aligned}
% \end{equation}
\begin{equation}\label{eq: DU_AU}
\begin{aligned}
D(U)(i,i) &= C_{\theta}(u_i,u_i) - C_{\theta}(u_i,\mbox{Pa}[u_i])C_{\theta}(\mbox{Pa}[u_i],\mbox{Pa}[u_i])^{-1}C_{\theta}(\mbox{Pa}[u_i],u_i)\;, \\
A(U)(i,\mbox{Pa}[u_{i}]) &= C_{\theta}(u_{i},\mbox{Pa}[u_{i}])C_{\theta}(\mbox{Pa}[u_{i}],\mbox{Pa}[u_{i}])^{-1}\;.
%\]
\end{aligned}
\end{equation}
Each row of $A(U)$ has exactly $m$ nonzero entries corresponding to the column indices in $\mbox{Pa}[u_{i}]$. The above structure implies that $y(s)$ and $y(s')$ are conditionally independent for any two points $s$ and $s'$ outside of $S$, given $y(S)$. The parameters $\theta$ will be estimated from the data $y(S)$ and predictions will be carried out using the conditional distribution of $y(U)$ given $y(S)$. In a Bayesian setting, $\theta$ will be sampled from its posterior distribution $p(\theta \given y(S))$,
% \begin{equation}\label{eq: nngp_response_posterior}
% \resizebox{0.9\hsize}{!}{$
% \begin{aligned}
% p(&\theta) \times \prod_{i=1}^n \frac 1{\sqrt{D(S)(i,i)}} \times \\ &\exp\left\lbrace -\frac{1}{2}z_{\theta}(S)^{\top}(I-A(S)^{\top})D(S)^{-1}(I-A(S))z_{\theta}(S) \right\rbrace\;,
% \end{aligned}$}
% \end{equation}
\begin{equation}\label{eq: nngp_response_posterior}
%\resizebox{0.9\hsize}{!}{$
\begin{aligned}
p(\theta) \times \left(\prod_{i=1}^n \frac 1{\sqrt{D(S)(i,i)}}\right) \times \exp\left\lbrace -\frac{1}{2}z_{\theta}(S)^{\top}(I-A(S)^{\top})D(S)^{-1}(I-A(S))z_{\theta}(S) \right\rbrace\;,
\end{aligned}
%$}
\end{equation}
where $z_{\theta}(S) = y(S) - m_{\theta}(S)$ and $p(\theta)$ is the prior distribution for $\theta$. 

Consider a specific example with the covariance function $C_{\theta}(s,s') = \sigma^{2}\exp(-\phi \|s-s'\|) + \tau^2\delta_{s=s'}$, where $\delta_{s=s'}$ is equal to one if $s=s'$ and $0$ otherwise, and $m_{\theta} = x(s)^{\top}\beta$ is a linear regression with spatial predictors $x(s)$ and corresponding slope vector $\beta$. Then $\theta = \{\beta, \sigma^2, \phi,\tau^2\}$ and one choice of priors could be
\begin{align*}
p(\theta) \propto U(\phi\given a_{\phi}, b_{\phi}) \times IG(\sigma^2\given a_{\sigma}, b_{\sigma}) \times IG(\tau^2\given a_\tau, b_\tau) \times N(\beta \given \mu_{\beta}, V_{\beta})\; ,
\end{align*}
where we are using standard notations for the above distributions as, e.g., in \citet{gelman2013}. The parameter space for this model is not high-dimensional and MCMC algorithms such as Gibbs sampling in conjunction with random-walk Metropolis (RWM) or %\textcolor{orange}
{Hamiltonian Monte Carlo (HMC)} can be easily implemented. Other approximate algorithms such as Variational Bayes or INLA can also be used. %\st{although subsequent results in this article use HMC.}

Once the parameter estimates (i.e., posterior samples) are obtained from (\ref{eq: nngp_response_posterior}) we can carry out predictive inference for $y(U)$ from the posterior predictive distribution
\begin{align}\label{eq: nngp_response_posterior_predictive}
p(y(U)\given y(S)) &= \int p(y(U)\given y(S),\theta)p(\theta\given y(S))d\theta %\\
%&
= \mbox{E}_{\theta\given y(S)}\left[N(y(U)\given \mu_{\theta}(U|\cdot), D(U))\right]\;,
\end{align}
where $p(y(U)\given y(S),\theta)$ is an $n'$-dimensional multivariate normal distribution with mean $\mu_{\theta}(U|\cdot) = m_{\theta}(U) + A(U)(y(S)-m_{\theta}(S))$ and conditional covariance matrix $D(U)$. Since $D(U)$ is diagonal, it is easy to sample from $p(y(U)\given y(S),\theta)$. For each $\theta$ sampled from (\ref{eq: nngp_response_posterior}), we sample an $n'$-dimensional vector $y(U)$ from $p(y(U)\given y(S),\theta)$. The resulting $y(U)$'s are samples from (\ref{eq: nngp_response_posterior_predictive}). The NNGP exploits the conditional independence between the elements of $y(U)$, given $y(S)$ and $\theta$, to achieve efficient posterior predictive sampling for $y(U)$. This assumption of conditional independence is not restrictive as the samples from (\ref{eq: nngp_response_posterior_predictive}) are not independent. In fact, the marginal covariance matrix of $y(U)$, given $\theta$ only, is $A(U)\tilde{C}_{\theta}(S,S)A(U)^{\top} + D(U)$, which is clearly not diagonal. 

\subsection{Latent NNGP model}\label{sec: nngp_latent}
\noindent Rather than model the outcome as an NNGP, as was done for the response model in the preceding subsection, one could use the NNGP as a prior for the latent process \citep{datta16}. In fact, as discussed in Section~4 of \citep{datta16}, the response model does not strictly follow the paradigm in (\ref{eq: generic_paradigm}) and it is not necessarily possible to carry out inference on a latent or residual spatial process after accounting for the mean.

A more general setting envisions a spatial regression model at any location $s$
\begin{equation}\label{eq: spatial_regression_model}
y(s) = m_{\theta}(s) + w(s) + \epsilon(s)\;,\quad \epsilon(s) \stackrel{iid}{\sim} N(0,\tau^2)\;,
\end{equation}
where, usually, $m_{\theta}(s) = x(s)^{\top}\beta$ and $w(s)$ is a latent spatial process capturing spatial dependence. Using definitions analogous to Section~\ref{sec: nngp_response}, we assume $\{w(s) : s\in D\} \sim NNGP(0,\tilde{C}_{\theta}(\cdot,\cdot))$, which means that for any $S$ and $U$, as constructed in (\ref{eq: nngp_linear_model}), $w \equiv w(S\cup U)$ will have a zero-centered multivariate normal law with covariance matrix $(I-A)^{-1}D(I-A)^{-\top}$. %An explicit covariance function for the NNGP is available and given by 
% \[
%  \tilde{C}_{\theta}(s,s') = \left\{\begin{array}{ll}
%  \tilde{C}_{\theta}(S)(i,j)  & \mbox{ if } s=s_i\in S,\; s'=s_j \in S \\
%  \tilde{C}_{\theta}(S)(i,Pa[s'])C_\theta(Pa[s'])^{-1}C_\theta(Pa[s'], s')  & \mbox{ if } s=s_i\in S,\; s' \notin S \\
%  \begin{array}{c}C_\theta(s, Pa[s])C_\theta(Pa[s])^{-1} \tilde{C}_{\theta}(S)(Pa[s],Pa[s'])\\ C_\theta(Pa[s'])^{-1}C_\theta(Pa[s'], s')
%  \end{array} &  \mbox{ if } s,s' \notin S 
%  \end{array} \right.
% \]
The posterior distribution to be sampled from is now given by
\begin{equation}\label{eq: nngp_latent_posterior}
%\resizebox{.9\hsize}{!}{$
\begin{aligned}
p(\theta) \times N(w \given &0, \tilde{C}_{\theta}(S,S)) %\\ 
\times \prod_{i = 1}^n N(y(s_i) \given m_{\theta}(s_i) + w(s_i), \tau^2) \;. %&
%\exp\left\lbrace -\frac{1}{2\tau^2}\sum_{i=1}^n(y(s_i) - m_{\theta}(s_i) - w(s_i))^2 \right\rbrace\;.
\end{aligned}
%$}
\end{equation}

It is easier to sample from (\ref{eq: nngp_response_posterior}) than from (\ref{eq: nngp_latent_posterior}) since the parameter space in the latter includes the high-dimensional random vector $w$ in addition to $\theta$. One option is to integrate out $w$ from (\ref{eq: nngp_latent_posterior}) which yields the posterior  
\begin{equation}\label{eq: nngp_latent_posterior_collapsed}
%\resizebox{.9\hsize}{!}{$
p(\theta) \times (\mbox{det}(\tilde{C}_{\theta}(S,S) + \tau^2 I_n))^{-\frac{1}{2}}\times \exp\left\lbrace -\frac{1}{2}\sum_{i=1}^n z_{\theta}(S)^{\top}\left(\tilde{C}_{\theta}(S,S) + \tau^2 I_n\right)^{-1}z_{\theta}(S) \right\rbrace\;,
%$}
\end{equation}
where $\mbox{det}(A)$ is the determinant of matrix $A$, $p(\theta)$ and $z_{\theta}(S)$ are as defined for (\ref{eq: nngp_response_posterior}). The parameter space has collapsed from $\{\theta,w\}$ to $\theta$, so (\ref{eq: nngp_latent_posterior_collapsed}) is called the collapsed version of (\ref{eq: nngp_latent_posterior}). %\sout{However, the matrix   $\left(\tilde{C}_{\theta}(S,S) + \tau^2 I\right)^{-1}$ is still high-dimensional, not sparse or computationally tractable as $\tilde{C}_{\theta}(S,S)^{-1}$. Efficient computations for obtaining \eqref{eq: nngp_latent_posterior_collapsed} will require sophisticated sparse-Cholesky decomposition in conjunction with parallel programming paradigms such as message-parsing-interfaces (MPI) or some variant thereof. Since our definition of ``modest'' does not extend to such environments, we will turn to conjugate models in the next section.} 
%\textcolor{orange}
{Efficient computations for obtaining \eqref{eq: nngp_latent_posterior_collapsed} requires a sparse-Cholesky decomposition for the large matrix $\left(\tilde{C}_{\theta}(S,S)^{-1} + \tau^{-2} I\right)$. This step can be complicated and expensive. To exacerbate the matter further, full Bayesian inference requires calculating the likelihood \eqref{eq: nngp_latent_posterior_collapsed} in each MCMC iteration as described in the algorithm of the ``collapsed'' model in Section~2.1 of \citet{finley2017applying}. To avoid such expenses, we turn to conjugate models in the next section.} 

\section{Conjugate Bayesian model} \label{sec: conj_Bayesian_model}
The response NNGP and latent NNGP models outlined in Sections~2.1~and~2.2%\ref{sec: nngp_response}~and~\ref{sec: nngp_latent}
, respectively, will still require iterative simulation methods such as MCMC for full Bayesian inference. Conjugate models, i.e., using conjugate priors, can provide exact Bayesian inference by exploiting analytic forms for the posterior distributions. While some specific assumptions are needed, these models are much faster to implement even for massive datasets. Here we develop conjugate NNGP models using the tractable Normal Inverse-Gamma (NIG) family of conjugate priors. We formulate a \emph{conjugate response model} (also formulated in \cite{finley2017applying} and is available in the \texttt{spNNGP} package from \url{cran.r-project.org/web/packages/spNNGP}) and a new conjugate latent NNGP model. These are conjugate versions of the models described in Sections~2.1~and~2.2%~\ref{sec: nngp_response}~and~\ref{sec: nngp_latent}
. We especially focus on the conjugate latent NNGP model and show how it can exploit sparsity by sampling from latent spatial processes over massive numbers of locations efficiently using a conjugate gradient algorithm for solving large sparse systems.  

\subsection{The NIG conjugate prior family}\label{sec: NIG_family} 
Let the spatial linear regression model be specified as
\begin{equation}\label{eq: latent_process_model}
y(S) = X\beta + w(S) + \epsilon(S)
\end{equation}
where $y(S)$, $w(S)$ and $\epsilon(S)$ are the realization of the corresponding processes defined in (\ref{eq: spatial_regression_model}) over the $n$ observed locations $S = \{s_1, \ldots, s_n\}$, $X$ is the $n \times p$ matrix of regressors with $i$-th row being a $1 \times p$ vector of regressors, $x(s_i)^{\top}$ at location $s_i \in S$. Henceforth, we suppress the dependence of $y$, $w$, $\epsilon$ and their covariance matrix on $S$ when this will not lead to confusion. Assume that $w \sim N(0, \sigma^2 C)$, $\epsilon \sim N(0, \delta^2 \sigma^2 I_n)$, where $C$ and $\delta^2 = \frac{\tau^2}{\sigma^2}$ are known. 
%One popular Bayesian model builds upon the regression of $y$ using conjugate priors by specifying a Normal-Inverse-Gamma(NIG) prior for $\{\beta, \sigma^2\}$. With the likelihood depending on the marginal distribution of $y$, the latent spatial random effects $w$ is integrated out and the parameter space collapses to $\{\beta, \sigma^2, \tau^2\}$.  A second model keeps $w$ in the parameter space by assigning a NIG prior for $\{\beta, w, \sigma^2\}$. To distinguish two conjugate models, we shall call the former as a conjugate response model and the latter as a conjugate random-effects model. Here we focus on the conjugate random-effects model. A conjugate random-effects model assigns an NIG prior for $\{\beta, w, \sigma^2\}$:
% \begin{equation}
%  \begin{array}{ll} p(\beta, w, \sigma^2)&
% = p(\beta | \sigma^2) p(w | \sigma^2) p(\sigma^2) 
% = N( \beta \given \mu_{\beta}, \sigma^2 V_{\beta}) N(w \given  0, \sigma^2 C) IG(\sigma^2 \given a_\sigma, b_\sigma)\\
% %&= ({1 \over \sigma^2})^{a + \frac{n+p}{2} + 1} exp( - \frac{1}{\sigma^2}\{b + \frac{1}{2}(\beta - \mu_{\beta})^\top V_\beta (\beta - \mu_{\beta}) + w^\top C^{-1}w  \}) 
% &\propto  ({1 \over \sigma^2})^{a_\sigma + \frac{n+p}{2} + 1} exp( - \frac{1}{\sigma^2} \{  
% b_\sigma + \frac{1}{2}(\gamma - \mu_{\gamma})^\top V_{\gamma}^{-1} (\gamma - \mu_{\gamma}) \}) \\
% &= p(\gamma, \sigma^2) = NIG(\mu_{\gamma}, V_{\gamma}, a_\sigma, b_\sigma)
% \end{array}	
% \end{equation}
%where 
Let $\gamma^\top = [\beta^\top, w^\top]$, $\mu_{\gamma}^\top = [\mu_{\beta}^\top, O^{\top}]$ and $V_\gamma = \left[\begin{array}{cc}
V_{\beta} & O \\ O & C \end{array} \right]$. The Normal-Inverse-Gamma (NIG) density yields a convenient conjugate prior,
\begin{equation}\label{eq: NIG_prior}
\begin{aligned}
p(\gamma,\sigma^2) &= NIG(\gamma,\sigma^2 \given \mu_{\gamma}, V_{\gamma}, a, b) %\\
%&
=  N(\gamma\given \mu_{\gamma},\sigma^2V_{\gamma}) \times IG(\sigma^2\given a,b) \;.
\end{aligned}
\end{equation}
The posterior distribution of the parameters, up to proportionality, is
\begin{equation}\label{eq: conj_post}
\begin{aligned}
p(\gamma, \sigma^2 \given y) \propto NIG(\gamma, &\sigma^2 \given \mu_\gamma, V_\gamma, a_\sigma, b_\sigma) %\\
\times N(y %&
\given [X : I_n] \gamma, \delta^2\sigma^2I_n)\;.
\end{aligned}
\end{equation}
The joint posterior distribution is of the form $NIG(\mu^*, V^*, a^*, b^*)$, where 
\begin{equation} \label{eq: NIG_post}
\begin{array}{ll}
y^* &= {1 \over \delta} y\;,\quad X^* = \left[{1 \over \delta}X, {1 \over \delta}I_n\right]\;, \\
\mu^* &= [V_{\gamma}^{-1} + X^{*\top} X^*]^{-1} (V_{\gamma}^{-1}\mu_{\gamma} + X^{*\top}y^*)\;, \\
V^* &= [V_{\gamma}^{-1} + X^{*\top} X^*]^{-1} \;,\\
a^*  &= a_\sigma + {n \over 2}\;, \\
b^*  &= b_\sigma + {1 \over 2} [\mu_{\gamma}^\top V_{\gamma}\mu_{\gamma} + y^{*\top} y^*  - \mu^{*\top}V^{*-1}\mu^*]\;.
\end{array}
\end{equation}
%\textcolor{blue}
{The prior of the regression coefficients $\beta$ is formulated as $N(\mu_\beta, V_\beta)$. The above model, however, also allows improper priors for $\beta$. When assigning improper priors for $\beta$, the precision matrix of the prior of $\gamma$ in \eqref{eq: NIG_post} becomes $V_{\gamma}^{-1} =  \left[\begin{array}{cc}
O & O \\ O & C^{-1} \end{array} \right]$, showing that no information from $\beta$`s prior contributes to the posterior distribution, and we can assume $\mu_{\gamma}^\top = [O^{\top}, O^{\top}]$ in \eqref{eq: NIG_post}.}
The marginal posterior distribution of $\sigma^2$ follows an $IG(a^*, b^*)$ and the marginal posterior distribution of $\gamma$ can be identified as a multivariate \textit{t}-distribution with mean $\mu^*$, variance $\frac{b^*}{a^*}V^*$ and degree of freedom $2a^*$(i.e. $\text{MVS-}t_{2a^*}(\mu^*, \frac{b^*}{a^*}V^*)$). Exact Bayesian inference is carried out by sampling directly from the joint posterior density: we sample $\sigma^2$ from $IG(a^*, b^*)$ and then, for each sampled $\sigma^2$, we draw $\gamma$ from its conditional posterior density $N(\mu^{\ast},\sigma^2V^{\ast})$. This yields posterior samples from (\ref{eq: conj_post}). Furthermore, note that once the posterior samples of $\sigma^2$ are obtained, we can obtain samples from $p(\tau^2\given y)$ by simply multiplying the sampled $\sigma^2$s with $\delta^2$. Thus, posterior samples are obtained without recourse to MCMC or other iterative algorithms.

\subsection{Conjugate response NNGP model}\label{se: conjugate_nngp_response}
\citet{finley2017applying} formulated a conjugate NNGP model for the response model described in Section~2.1%\ref{sec: nngp_response}
. This is formed by integrating out $w(S)$ from (\ref{eq: latent_process_model}) and applying an NNGP approximation to the marginal covariance matrix of $y(S)$. The model can be cast as a conjugate Bayesian linear regression model
\begin{equation}\label{eq: conj_post_nngp_response}
%\resizebox{.88\hsize}{!}{$
p(\beta, \sigma^2 \given y) \propto NIG(\beta, \sigma^2 \given \mu_\beta, V_\beta, a_\sigma, b_\sigma) \times N(y \given X \beta, \sigma^2\tilde{K})\;,%$}
\end{equation}
where $\tilde{K}$ is the NNGP approximation of $K = C + \delta^2I$, $C$ and $\delta^2$ are as described in Section~\ref{sec: NIG_family}. Also, $\tilde{K}^{-1} = \sigma^2(I-A(S)^{\top})D(S)^{-1}(I-A(S))$ with $A(S)$ and $D(S)$ as described in Section~\ref{sec: nngp_response}. We will refer to (\ref{eq: conj_post_nngp_response}) as the conjugate response NNGP model. Note that this model can estimate $\{\beta,\sigma^2\}$ and also impute the outcome at unknown locations, but does not permit inference on the latent process $w(\cdot)$. %\textcolor{blue}
{The reason why a conjugate response NNGP model cannot provide inference on the latent process is that the construction of the response NNGP will not guarantee the existence of a well-defined latent process. It is pointed out in Section~4 of \citep{datta16} that the eigenvalue of $\tilde{K}$ may be less than $\delta^2$, consequently the covariance matrix of the posterior distribution of $w$ need not be positive definite for every proper $\delta^2$, $\mu_\beta$ and $V_\beta$.} We address this shortcoming with a new conjugate latent NNGP model in the next section.

\subsection{Conjugate latent NNGP model}\label{sec: conjugate_nngp}
The conjugate models in Section~\ref{sec: NIG_family} 
works for any covariance matrix $C$. Here, we derive a conjugate latent NNGP model that will subsume inference on $w(\cdot)$. We rewrite the covariance matrix $\tilde{C}_\theta(S,S)$ in section~\ref{sec: nngp_latent} for $w(S)$ as $\sigma^2\tilde{M}_\phi$ with fixed parameter $\phi$. Note that $\tilde{M}_\phi$ is the NNGP approximation of the dense matrix $M$, where $C = \sigma^2 M$. Specifically, $\tilde{M}_\phi^{-1} = (I - A_M)^{\top}D_M^{-1}(I - A_M)$, where $A_M$ and $D_M$ depend only on $\phi$. We recast the model as
\begin{equation} \label{eq:conjugate_sparse}
\begin{array}{c}
\underbrace{ \left[ \begin{array}{c} {1 \over \delta} y\\ L_\beta^{-1} \mu_\beta \\ 0 \end{array} \right]}
= \underbrace{ \left[ \begin{array}{cc} {1 \over \delta}X& {1 \over \delta}I_n \\ L_\beta^{-1}& O \\  O& D_M^{-{1 \over 2}}(I-A_M) \end{array} \right] }
\underbrace{ \left[ \begin{array}{c} \beta \\ w \end{array} \right]}+ \underbrace{ \left[ \begin{array}{c} \eta_1 \\ \eta_2 \\ \eta_3 \end{array} \right]}\\
\hspace{0.2cm} y_{*} \hspace{0.5cm} = \hspace{1.4cm}X_{*} \hspace{1.8cm}\gamma \hspace{0.4cm}+ \hspace{0.4cm}\eta \hspace{2.4cm}
\end{array}
\end{equation}
where $L_\beta$ is the Cholesky decomposition of the $p \times p$ matrix $V_\beta$, and $\eta \sim N(0, \sigma^2 I_{2n + p})$. The joint posterior distribution of $\gamma$ and $\sigma^2$ follows an NIG distribution
\begin{equation}\label{eq: joint_posterior_cojugate}
p(\gamma, \sigma^2 \given y) = NIG(\gamma, \sigma^2 \given  \hat{\gamma}, (X_{*}^{\top}X_{*})^{-1}, a_*, b_* )
% \propto IG(\sigma^2 | a_*, b_*) \times N(\gamma \given \hat{\gamma}, \sigma^2(X_{*}^{\top}X_{*})^{-1}) \\
\end{equation}
where $\displaystyle \hat{\gamma} = (X_{*}^{\top}X_{*})^{-1}   X_{*}^{\top} y_{*}$, $\displaystyle a_{*} =  a_\sigma + {n \over 2}$ and $\displaystyle b_{*} = b_\sigma +  {1 \over 2}(y_{*} - X_{*}\hat{\gamma})^{\top}(y_{*} - X_{*}\hat{\gamma})$.
% all the parameters are available in terms of the sparse matrix $X_{*}$, vector $y_{*}$ and $a_\sigma$ and $b_\sigma$:
% \begin{equation} \label{eq: conjugate_posterior_parameters}
% \begin{array}{ll} 
% \hat{\gamma} &=  (X_{*}^{\top}X_{*})^{-1}   X_{*}^{\top} y_{*} \\
% a_{*}& =  a_\sigma + {n \over 2} \\
% b_{*}&= b_\sigma +  {1 \over 2}(y_{*} - X_{*}\hat{\gamma})^{\top}(y_{*} - X_{*}\hat{\gamma})
% \end{array}
% \end{equation}
Evaluating the posterior mean of $\gamma$ involves solving $X_{*}^{\top}X_{*} \hat{\gamma} = X_{*}^{\top} y_{*}$, which requires $\mathcal{O}({1 \over 3}(n+p)^3)$ flops.  However, when $p \ll n$, the structure of $X_{*}$ ensures low storage complexity. Also, $X_{*}^{\top}X_{*} =$
\begin{equation}
%\resizebox{.88\hsize}{!}{$
\begin{aligned}
% X_{*}^{\top}X_{*} =\\
\left[ \begin{array}{cc} {1 \over \delta^2} X^\top X + 
L_\beta^{-\top} L_\beta^{-1} & {1 \over \delta^2} X^\top \\
{1 \over \delta^2} X & {1 \over \delta^2} I_n + 
(I_n - A_M)^\top D_M^{-1} (I_n - A_M)\end{array} \right]
\end{aligned}
%$}
\end{equation}
Since $(I_n - A_M)$ has less than $n(m+1)$ nonzero elements and each of its row has at most $m+1$ nonzero elements, the storage of the $n \times n$ matrix $(I_n - A_M)^\top D_M^{-1} (I_n - A_M)$ is less than $n(m+1)^2$, and the computational complexity is less than $nm + n(m+1)^2$. 
%The non-zero entries in $X_{*}^{\top}X_{*}$ is less than $p^2 + 2np + n(m+1)^2$. 

This sparsity in $X_{*}^{\top}X_{*}$ can be exploited by a conjugate gradient %\textcolor{blue}
{(CG) method} \citep[see, e.g.,][]{golub2012}. %which transforms the problem of solving a linear system into minimizing a quadratic function.
%\textcolor{blue}
{CG is an iterative method for solving $Ax = b$ when $A$ is a symmetric positive definite matrix. The underlying idea is to recognize that a solution of the linear system $Ax = b$ minimizes the quadratic function $\phi(x) = \frac{1}{2}x^\top A x - x^\top b$. CG is an iterative procedure that generates a sequence of approximate solutions $\{x_{k}\}_{k = 1, 2, \ldots}$ that converges to $x = A^{-1}b$ in at most $n$ iterations. Briefly, the procedure starts with an initial value $x_0$ and setting $r_0 = b-Ax_0$ and $q_0 = r_0$. Then, at the $k+1$-th iteration we compute the following three quantities for each $k=0,1,2,\ldots$: (i) $x_{k+1} = x_k + \frac{\|r_k\|^2}{q_{k+1}^{\top}Aq_{k+1}}q_{k+1}$; (ii) $r_{k+1} = b - Ax_{k+1}$; and (iii) $q_{k+1} = r_{k+1} + \left(\frac{\|r_{k+1}\|}{\|r_{k}\|}\right)^2q_{k}\;$.
The matrix $A$ is involved only in matrix-vector multiplications. Due to the sparsity of $A$, the computational cost per iteration is $\mathcal{O}(n)$ flops. The sparsity in $A$ also implies that CG is more memory efficient than direct methods such as the Cholesky decomposition. %Iteration $k$ of the CG algorithm returns $x_k$ as the minimizer of the target quadratic function $\phi(x)$ over a subspace $SP_k$ of dimension $k$. The next iteration produces minimizer $x_{k+1}$ over the space spanned by the $SP_k$ and the gradient over the target quadratic function at $x_{k}$, making the algorithm converge in super-linear time.
%The quadratic function with approximate solutions $\{\phi(x_k)\}_{k = 0, 1, \ldots}$ is monotonically decreasing. 
A sufficiently good approximation is often obtained %\textcolor{blue}{\sout{in a few iterations} 
in iterations much less than $n$ \citep{banerjee2014linear}, hence the performance of the conjugate gradient algorithm will be competitive when $n$ is large.} This enables posterior sampling of the latent process $w(S)$ in high-dimensional settings. The algorithm for sampling $\{\gamma, \sigma^2\}$ from (\ref{eq: joint_posterior_cojugate}) using the conjugate gradient method is given below.
%\subsection{Recovering Latent Spatial Effects Using Conjugate Gradient}\label{sec: conjugate_gradient}
%The following Algorithm illustrates the procedures of how to sample $\{\gamma, \sigma^2\}$ from~(\ref{eq: joint_posterior_cojugate}):

\noindent{\rule{0.99\textwidth}{1pt}\\
{\fontsize{8}{8}\selectfont
	\textbf{Algorithm~1}: 
	Sample $\{\gamma, \sigma^2\}$ from conjugate latent NNGP model\\ %with fixed $\phi$ and $\delta^2$ \\
	[-2pt]\rule{0.99\textwidth}{1pt}\\[-12pt]
	\begin{enumerate}
		\item[1.] Fixing $\phi$ and $\delta^2$, obtain $L_\beta^{-1}\mu_\beta$ and $L_\beta^{-1}$: 
		\begin{itemize}
			\item Compute a Cholesky decomposition of $V_\beta$ to get $L_\beta$ \hfill{$\mathcal{O}(p^3)$}
			\item Compute $L_\beta^{-1}$ and $L_\beta^{-1} \mu_\beta$    \hfill{$\mathcal{O}(p^2)$}
		\end{itemize}
		\item[2.] Obtain the posterior mean for $\gamma$:
		\begin{itemize}
			\item Construct $A_M$ and $D_M$ as described, for example, in \citet{finley2017applying} \hfill{$\mathcal{O}(nm^3)$}
			\item Construct $X_{*}$ and $Y_{*}$ from (\ref{eq:conjugate_sparse})
			\hfill{$\mathcal{O}(nm)$}
			\item Calculate $X_{*}^{\top}X_{*}$ and $X_{*}^{\top}y_{*}$ \hfill{$\mathcal{O}(n(m+1)^2)$}
			\item Use conjugate gradient to solve $X_{*}^{\top} X_{*}\hat{\gamma} = X_{*}^{\top} y_{*}$ 
		\end{itemize}
		\item[3.] Obtain posterior samples of $\sigma^2$ 
		\begin{itemize}
			\item Calculate $a_{*}$ and $b_{*}$ as given below (\ref{eq: joint_posterior_cojugate}) \hfill{$\mathcal{O}(n(m+4+p))$}
			\item Sample $\sigma^2$ from $IG(a_{*}, b_{*})$
		\end{itemize}
		\item[4.] Obtain posterior samples of $\gamma$
		\begin{itemize}
			\item Generate $u \sim N(0, \sigma^2 I_{2n + p})$
			\item Calculate $v$ by solving $X_{*}^{\top} X_{*} v = X_{*}^{\top}u$ using conjugate gradient
			\item Obtain $\gamma = \hat{\gamma} + v$  \hfill{$\mathcal{O}(n)$}
		\end{itemize}
	\end{enumerate}	
	\vspace*{-8pt}
	\rule{0.99\textwidth}{1pt}
} }

It is readily seen that the $v$ in step~4 follows a Gaussian distribution with variance $\sigma^2 (X_{*}^{\top} X_{*})^{-1}$. Note that Algorithm~1 implements the conjugate gradient method for an $n + p$-dimensional linear system in steps~2~and~4. Since $X_{*}$ and $y_{*}$ depend only on $\{\phi, \delta^2\}$, the linear equation in step~2 only need to be solved once for each choice of $\{\phi, \delta^2\}$.% and the conjugate gradient in step~4 needs to be repeated for each sample of $\nu$. 

{The main contribution of the conjugate gradient method lies in obtaining the posterior estimator $\hat{\gamma}$ (step 2) and generating samples from a high dimensional Gaussian distribution (step 4). It is worth pointing out that the conjugate gradient method does not easily produce the determinant of a large matrix. Hence, a sparse Cholesky decomposition is still unavoidable for the collapsed NNGP model formulated in equation \eqref{eq: nngp_latent_posterior_collapsed}, where $\det\left(\tilde{C}_{\theta}(S,S) + \tau^2 I\right)$ changes with the hyper-parameters $\theta$.}

\subsection{Posterior predictive inference for conjugate latent NNGP }
We extend the predictive inference for the response NNGP model in Section~2.1 to the conjugate latent NNGP model. Assume $w(U)$ and $y(U)$ are the realization of the latent process and the response process over the $n'$ locations $U = \{ u_1, \ldots, u_{n'} \}$ where we wish to predict. let $C_\theta(\cdot, \cdot)$ be the covariance function for the latent process $w(s)$ in \eqref{eq: spatial_regression_model}, $Pa[u_i]$ be the nearest neighbors of $i$th location in $U$ as defined in section 2.1. Define $A_u = A(U)$ and $D_u = \frac{1}{\sigma^2}D(U)$ where $A(U)$ and $D(U)$ are constructed by \eqref{eq: DU_AU}. %be an $n' \times n$ sparse matrix whose $i$ th row has nonzero entries in the positions indexed by $Pa[u_i]$
%\begin{equation}
%A_u[i, Pa[u_i]] = C_\theta(u_i, Pa[u_i]) C_\theta(Pa[u_i], Pa[u_i])^{-1}
%\end{equation}
%$D_u$ be an $n' \times n'$ diagonal matrix where
%\begin{equation}
%D_u[i, i] = 1 - \frac{1}{\sigma^2}C_\theta(u_i, Pa[u_i]) C_\theta(Pa[u_i], Pa[u_i])^{-1} C_\theta(Pa[u_i], u_i).
%\end{equation}
Here, $\sigma^2$ refers to the variance of the latent process $w(s)$, and $A_u$ and $D_u$ are defined in the way that they only depend on fixed parameter $\phi$. According to the definition of NNGP process over the whole domain given in section~2%\ref{sec: nngp_review}
, the joint distribution of $w(U)$ and $\gamma, \sigma^2$ given $y(S)$ follows:
\begin{equation}\label{eq: joint_w(US)_gamma_Conj_LNNGP}
\begin{aligned}
p(w(U), \gamma, \sigma^2 \given y(S)) =  N(w(U) \given [O:A_u] \gamma, \sigma^2 D_u) \times NIG (\gamma, \sigma^2 \given \hat{\gamma}, (X_*^\top X_*)^{-1}, a_*, b_*) 
\end{aligned}
\end{equation}
Marginalizing the joint distribution \eqref{eq: joint_w(US)_gamma_Conj_LNNGP} over $\gamma$ and $\sigma^2$, the posterior distribution of $w(U)$ can be identified as a multivariate \textit{t}-distribution:
\begin{equation}
	w(U) \given y(S) \sim \text{MVS-}t_{2a_*} \left(\mu_{wu}, \frac{b_*}{a_*}V_{wu}\right)
\end{equation}
%\begin{equation}\label{eq: w(U)_sigma_NIG}
%p(w(U), \sigma^2 \given y(S)) = NIG(w(U), \sigma^2 \given \mu_{wu}, V_{wu}, a_*, b_*)
%\end{equation}
where
\[
\begin{aligned}
\mu_{wu} = [O: A_u] \hat{\gamma} \;, \;
V_{wu} = [O: A_u] (X_*^\top X_*)^{-1} \begin{bmatrix} O\\  A_u^\top\end{bmatrix}+ D_u \; .
\end{aligned}
\]
It is straightforward to see that the joint posterior distribution of $\{y(U), w(U), \gamma, \sigma^2\}$ is
\begin{equation}\label{eq: y(U)_w(U)_gamma_sigma_Conj_LNNGP}
%\resizebox{0.88\hsize}{!}{$
\begin{aligned}
&p(y(U), w(U), \gamma, \sigma^2 \given y(S))%\\
%&
= N(y(U) \given X(U)\beta + w(U), \sigma^2 \delta^2 I_{n'}) \times p(w(U), \gamma, \sigma^2 \given y(S))\;,
\end{aligned}
%$}
\end{equation}
which is the product of the conditional distribution of $y(U)$ from the spatial linear regression model \eqref{eq: spatial_regression_model} and the posterior distribution \eqref{eq: joint_w(US)_gamma_Conj_LNNGP}. It can be shown that the posterior distribution of the predictive process $y(U)$ and $\sigma^2$ follows an NIG after marginalizing out $\gamma$ and $w(U)$, and the posterior distribution of $y(U)$ follows a multivariate \textit{t}-distribution:
\begin{equation} \label{eq: post_y(U)_conj_LNNGP}
 y(U) \given y(S) \sim  \text{MVS-}t_{2a_*} \left(\mu_{yu}, \frac{b_*}{a_*}V_{yu}\right)
\end{equation}
where
\[
\begin{aligned}
\mu_{yu} &= [X(U): A_u]\hat{\gamma}\; \mbox{ and }\; %, \\
V_{yu} %&
= [X(U): A_u] (X_*^\top X_*)^{-1}  \begin{bmatrix} X(U)^\top\\ A_u^\top \end{bmatrix} + \delta^2I_{n'} + D_u\; .
\end{aligned}
\]
%where $\mu_{yu} = [X(U): A_u]\hat{\gamma}$ and $V_{yu} = [X(U): A_u] (X_*^\top X_*)^{-1}  \begin{bmatrix} X(U)^\top\\ A_u^\top \end{bmatrix} + \delta^2I_{n'} + D_u$

Sampling $w(U)$ $y(U)$ from their posterior distribution requires taking Cholesky decomposition of matrix $V_{wu}$ and $V_{yu}$. Since the matrix $(X_*^\top X_*)^{-1}$ is involved in the calculation, the required computation power is expensive and the calculation quickly become forbidden when the number of locations to predict is large. Rather than direct sampling, we recommend using a two stage sampling method based on the joint distribution \eqref{eq: joint_w(US)_gamma_Conj_LNNGP} and \eqref{eq: y(U)_w(U)_gamma_sigma_Conj_LNNGP} in this subsection. First, obtain the posterior samples $\{\gamma^{(l)}, \sigma^{2(l)}\}_{l = 1}^L$. Then generate the posterior samples of $w(U)$ through $w(U)^{(l)} \sim N( [O:A_u] \gamma^{(l)}, \sigma^{2(l)} D_u)$ for $l = 1, \ldots, L$. Finally use $y(U)^{(l)} \sim N(X(U)\beta^{(l)} + w(U)^{(l)}, \delta^2\sigma^{2(l)})$ to generate the posterior samples of $y(U)$.

%\begin{equation}
%\begin{aligned}
%& = N(\begin{pmatrix}w\\w(U)\end{pmatrix} \given 0, \sigma^2 \begin{bmatrix}(I_n - A_M) & 0\\ -C_u& I_{n'} \end{bmatrix} ^{-1} \begin{bmatrix} D_M & 0 \\ 0 & I_{n'} - D_u \end{bmatrix} \begin{bmatrix}(I_n - A_M)^\top & -C_u^\top\\ 0 & I_{n'}\end{bmatrix})  \\
%&\times N( \begin{pmatrix}y(S)\\y(U) \end{pmatrix} \given \begin{bmatrix}X\\ X_u \end{bmatrix}\beta + \begin{bmatrix}w(S)\\ w(U) \end{bmatrix} , \sigma^2 \delta^2I_{n + n'}) \times NIG(\beta, \sigma^2 \given \mu_\beta, V_\beta, a_\sigma, b_\sigma) \\
%& \propto N(y(U) \given X_u \beta + c_u w, \sigma^2\delta^2I_{n'}) \times N(w(U) \given C_u w, \sigma^2 (I_{n'} - D_u)) \times NIG(\gamma, \sigma^2 \given \hat{\gamma}, (X_{*}^{\top}X_{*})^{-1}, a_*, b_*)
%\end{aligned} 
%\end{equation}

\subsection{Inference of $\phi$ and  $\delta^2$}\label{sec: inference_fixed_parameters}

Algorithm~1 provides the exact posterior sampling of the process parameters after specifying $\phi$ and $\delta^2$. This motivates us to estimate all the process parameters by first obtaining the inference of a small set of parameters $\phi$ and $\delta^2$, then implement\textcolor{blue}{ing} Algorithm~1 to sample $\{\gamma, \sigma^2\}$. 
%When we sample from $p(\phi, \delta^2 \given y)$, the generated samples of $\{\gamma, \sigma^2\}$ follow the marginal posterior distribution $p(\gamma, \sigma^2 \given y)$; 
When we fix $\phi$ and $\delta^2$ at a point estimator (i.e. $\arg\max{\{p(\phi, \delta^2 \given y)\}}$), the conjugate latent NNGP model becomes a special case of fitting latent NNGP model with Empirical Bayes method.

Here we propose a $K$-folder cross-validation algorithm for picking a point estimate of $\{\phi,\delta^2\}$ of the conjugate Latent NNGP. We first split the data randomly into K folds and denote the $k$-th folder of the observed locations $S[k]$, whereas $S[-k]$ denotes the observed locations without $S[k]$. Then we fit the predictive mean $E[y(S[k]) \given y(S[-k])]$ by the posterior distribution given in  \eqref{eq: post_y(U)_conj_LNNGP}. We use the Root Mean Square Predictive Error (RMSPE)(\citet{yeniay2002comparison}) to select $\phi$ and $\delta^2$ from a gird of candidate values. %\textcolor{blue}
{The initial candidates for $\{\phi, \delta^2\}$ comes from a coarse grid. The range of the grid is decided based on interpretation of the hyper-parameters. Specifically, the spatial decay $\phi$ describes how the spatial correlation decreases as the distance between two locations increases. Define $\mbox{maxdist}(S) := \mbox{max}_{s, t \in S}\{d(s, t)\}$ where $d(s, t)$ is the distance between location $s$ and $t$. The lower bound of the candidate value of $\phi$ is set at $\frac{3}{\mbox{maxdist}(S)}$, which indicates that the spatial correlation drops below 0.05 when the distance reaches $\mbox{maxdist}(S)$. The upper bound can be initially set as 100 times of the lower bound $\frac{300}{\mbox{maxdist}(S)}$. For $\delta^2$, we need to use reasonable assumptions on the variance components. %When the ratio of the noise variance to the spatial variance $\delta^2$ falls out of $[0.001, 1000]$, the spatial process will dominate or be dominated by the noise process. Hence
A suggested wide range for $\delta^2$ can be $[0.001, 1000]$, which accommodates one variance component substantially dominating the other in either direction. The prior information from the related studies of the data as well as the estimators from the variogram also provide the candidate value of $\{ \phi, \delta^2 \}$. Functions like \texttt{variofit} in the R package \texttt{geoR} \citep{geoR} can provide empirical estimates for $\{\phi, \delta^2\}$ from an empirical variogram. After initial fitting, we can shrink the range and refine the grid of the candidate values for more precise estimators.} Algorithm~2 describes K-fold cross-validation for choosing $\phi$, $\delta^2$ in the conjugate latent NNGP model.

\noindent{ \rule{0.99\textwidth}{1pt} \\
	{\fontsize{8}{8}\selectfont
	\textbf{Algorithm~2}: Cross-validation of tuning $\phi$, $\delta^2$ for conjugate latent NNGP model\\
	[-2pt]\rule{0.99\textwidth}{1pt}\\[-12pt]
	\begin{enumerate}
		\item[1.] Split the data into $K$ folds, and build neighbor index.
		\begin{itemize}
			\item Build nearest neighbors for $S[-k]$
			\item Find the collection of nearest neighbor set for $S[k]$ among $S[-k]$.
		\end{itemize} 
		\item[2.] Fix $\phi$ and $\delta^2$, Obtain the posterior mean for $\gamma_k= \{ \beta, w(S[-k]) \}$ after removing the $k^{th}$ fold of the data:
		\begin{itemize}
			\item Use step 1-2 in Algorithm~1 to obtain $\hat{\gamma}_k$ 
		\end{itemize}
		\item[3.] Predicting posterior means of $y(S[k])$
		\begin{itemize}
			\item Construct matrix $A_u$ for $S[k]$ 
			\item According to \eqref{eq: post_y(U)_conj_LNNGP}, the predicted posterior mean follows\\
			$\hat{y}(S[k]) = E[y(S[k]) \given y(S[-k])] = [X(U): A_u] \hat{\gamma}$
		\end{itemize}
		\item[4.] Root Mean Square Predictive Error (RMSPE) over K folds
		\begin{itemize}
			\item Initialize $e = 0$\\
			\text{ } for ($k$ in $1:K$)  \\
			\text{ } \quad \quad for ($s_i$ in $S[k]$)\\
			\text{ } \quad \quad \quad \quad $ e =e  + (y(s_i) - \hat{y}(s_i))^2 $
		\end{itemize}
		\item[5.] Cross validation for choosing $\phi$ and $\delta^2$
		\begin{itemize}
			\item Repeat steps (2) - (4) for all candidate values of $\phi$ and $\delta^2$
			\item Choose $\phi_0$ and $\delta_0$ as the value that minimizes the average RMSPE
		\end{itemize}
	\end{enumerate}	
	\vspace*{-8pt}
	\rule{0.99\textwidth}{1pt}
}}

The main computational burden lies in step 1 in Algorithm~2. However, step 1 serves as a pre-calculation for the whole cross-validation since it only need to be calculated for once. We recommend using a KD-tree algorithm provided in R package \texttt{spNNGP} \citep{spNNGP} to build the nearest neighbor matrics. Step 2 dominates the computational requirement in Algorithm~2 after the pre-calculation, which calls Algorithm 1 for $k$ times for each choice of $\{ \phi, \delta^2 \}$. 

An alternative approach for choosing point estimates of $\{\phi,\delta^2\}$ is to carry out the cross-validation with the conjugate response NNGP model in (\ref{eq: conj_post_nngp_response}). The practical advantage here is that the function \texttt{spConjNNGP} within the \texttt{spNNGP} package in \texttt{R} can be used to carry out the cross-validation. The algorithm behind \texttt{spConjNNGP} is exactly linear in $n$ and highly efficient in its implementation. Empirical studies reveal that the response NNGP model and the latent NNGP model provide similar optimal choices for $\{\phi, \delta^2\}$ when using the K- folder cross-validation. %Empirical studies reveal that response NNGP performs similarly with the latent process model \textcolor{orange}{for referring $\{\phi, \delta^2\}$. %\sout{While through a Directed Acyclic Graph (DAG) built on an augmented latent space(parameter space including latent process),  \citep{katzfuss2017general} show that the Kullback-Leibler divergence(KL-D) \citep{gneiting2007strictly} from the latent NNGP model to the full GP model is no more than that of the response NNGP model, which suggests that the latent NNGP model is better than the response NNGP model. This raises the concern of using response NNGP model instead of latent NNGP model for model fitting.  However, the claim in \citep{katzfuss2017general} based on KL-Ds on an augmented space (parameter space with latent process $w$) is not guaranteed to hold on a collapsed space (parameter space without latent process $w$). We provide an example to discuss this claim in the next subsection.}}

\section{Simulation Study}\label{sec: simulation_study}

We use a simulation study in this section to discuss the performance of the aforementioned models in Sections~\ref{sec: nngp_review}~and~\ref{sec: conj_Bayesian_model}. Algorithm~1 were programmed in R which calls the \texttt{Rstan} environment \citep{rstan} for building matrix $A_M$ and $D_M$. The conjugate gradient solver for sparse linear systems was implemented through \texttt{RcppEigen} \citep{RcppEigen}, which calls a Jacobi preconditioner \citep[see, e.g., page 653 in][]{golub2012} by default. We provide a brief discussion on preconditioned conjugate gradient algorithms in Section~\ref{sec: conclusion}.
%, and the program for constructing $A_M$ and $D_M$ defined in Section~\ref{sec: nngp_review} are written in Rstan. 
The nearest-neighbor sets were built using the \texttt{spConjNNGP} function in the \texttt{spNNGP} package. All simulations were conducted on a OS High sierra system (version 10.13.4) with 16GB RAM and one 3.1 GHz Intel-Core i7 processors.

\subsection{Univariate simulation study} \label{sec: sim_uni}
We generated data using the spatial regression model in (\ref{eq: spatial_regression_model}) over a set of $n=1200$ spatial locations within a unit square. The true values of the parameters generating the data are supplied in Table~\ref{table:sim}. The size of the data set was kept moderate to permit comparisons with the expensive full GP models. The model had an intercept and a single predictor $x(s)$ generated from a standard normal distribution. An exponential covariance function was used to generate the data.  
% response $Y(s)$ along with a predictor $x(s)$ at $n = 1200$ randomly generated locations within a unit square domain by the following model:
% \begin{equation}
% y(s) = \beta_0 + x(s)\beta_1 + w(s) + \epsilon(s), \hspace{1cm} \epsilon(s) \sim N(0, \tau^2)
% \end{equation}
% where the zero-centered spatial random effect $w(s)$ were sampled from a Gaussian process with a covariance function $C_\theta$ specified by exponential:
% \begin{equation} \label{exp_K}
% C_\theta(s_i, s_j) = \sigma^2\exp(-\phi||s_i-s_j||), \hspace{0.5cm} s_i,s_j \in S \hspace{0.5cm} %\theta = \{ \sigma^2, \phi \}
% \end{equation} 
%Parameters value are listed in the first column of Table 1. The predictor $x$ were generated from $N(0,1)$.  

Candidate models for fitting the data included full Gaussian process based model (labeled as full GP in Table~\ref{table:sim}), a latent NNGP model with $m = 10$ neighbors and a conjugate latent NNGP model with $m = 10$ neighbors. These models were trained using $n=1000$ of the $1200$ observed locations. And the remaining $200$ observations were withheld to assess predictive performance. The full Gaussian process based model was implemented with function \textit{spLM} in R package \textit{spBayes}. The latent NNGP model was conducted with function \textit{spNNGP} in R package \textit{spNNGP}. The fixed parameters$\{ \phi, \delta^2 \}$ for the conjugate latent NNGP model were picked through the $k$-th folder cross-validation algorithm (Algorithm~2). And the choice from \texttt{spConjNNGP} coincide with the cross-validation for the conjugate latent NNGP model. 
%In this study, we generate one sample of $\{\gamma, \sigma^2\}$ for each iteration of MCMC chains for the response model after warm-up, which yields 3000 samples in total.  
%The model where $\phi$ and $\delta^2$ are fixed is referred to as \textbf{Model2}. The values of $\{\phi, \delta^2\}$ are fixed by finding the minimum RMSPE using the conjugate response NNGP model in (\ref{eq: conj_post_nngp_response}) based upon the training data. 

The intercept and slope parameters $\beta$ were assigned improper flat priors. The spatial decay $\phi$ was modeled using a fairly wide uniform prior $U(2.2, 220)$. We use Inverse-Gamma priors $IG(2,b)$ (mean $b$) for the nugget ($\tau^2$) and the partial sill ($\sigma^2$) in order to compare the conjugate Bayesian models with other models. The shape parameter was fixed at $2$ and the scale parameter was set from the empirical estimate provided by the variogram using the \texttt{geoR} package \citep{geoR}. The parameter estimates and performance metrics are provided in Table~\ref{table:sim}. 
\begin{table} 
	\centering
	\def\~{\hphantom{0}}
	\caption{Simulation study summary table: posterior mean (2.5\%, 97.5\%) percentiles} \label{table:sim}
	{\begin{tabular*}{\textwidth}
			{@{}c@{\extracolsep{\fill}}c@{\extracolsep{\fill}} c@{\extracolsep{\fill}}c@{\extracolsep{\fill}}c@{\extracolsep{\fill}}
				c@{\extracolsep{\fill}}c@{\extracolsep{\fill}}@{}}
			\hline
		%	&&\phantom{ab}& \multicolumn{2}{c} {Response} & \phantom{ab}&\multicolumn{2}{c}{Latent} \\
			%[1pt] \cline{4-5} \cline{7-8} \\ [-10pt]
			& True && Full GP  && NNGP & Conj LNNGP \\
			\hline
			$\beta_0$ & 1 	&&  1.07(0.72, 1.42) &&  1.10 (0.74, 1.43)  &  1.06 (0.76, 1.46)	  \\
			$\beta_1$ & -5 	&&   -4.97 (-5.02, -4.91) &&  -4.97 (-5.02, -4.91)  &	-4.97 (-5.02, -4.91) \\
			$\sigma^2$ & 2 &&  1.94 (1.63, 2.42) &&  1.95 (1.63, 2.41)  &   1.94 (1.77, 2.12) \\
			$\tau^2$ & 0.2 &&  0.14 (0.07, 0.23)  && 0.15 (0.06, 0.24)   & 0.17 (0.16, 0.19)\\
			$\phi$ & 16 && 19.00 (13.92, 23.66)  &&   18.53 (14.12, 24.17)  &  17.65 \\ 
			\hline
%			KL-D &--&& 1048.1(932.85, 1930.38)   &&  1045.61(634.20, 2321.00) & 795.37(726.18, 876.96) \\ %(mean, 2.5\%, 97.5\%)
			KL-D &--&& 4.45(1.16, 9.95)   &&  5.13(1.66, 11.39) & 3.58(1.27, 8.56) \\ 
			MSE(w) &--&&  297.45(231.62, 444.79 ) && 303.38(228.18, 429.54)  &  313.28 (258.96, 483.75)\\
			RMSPE &--& &  0.94  && 0.94  &  0.94 \\
			time(s) & -- & & 2499 + 23147   &&  109.5   &  12 + 0.6 \\
			\hline
		\end{tabular*}}
		\vspace*{-6pt}
\end{table}
The summaries for the full Gaussian process based model and the latent NNGP model were based on 1 MCMC chain with $20,000$ iterations. The number of iterations was taken to be large enough to guarantee the convergence of the MCMC chains. We took the first half of the MCMC chains as burn-in. The inference from the conjugate latent NNGP model were based on 300 samples. 300 samples is sufficient for the conjugate latent NNGP model since the conjugate model provides independent samples from the exact posterior distribution. We don't need extra memory for burn-in, and the samples from the conjugate model are more efficient than that from MCMC algorithms.

All models were assessed by the Kullback-Leibler divergence (labeled KL-D;  \citet{gneiting2007strictly}) and the out-of-sample root mean squared prediction error (RMSPE) (\citet{yeniay2002comparison}). The KL-D between true distribution $Q$ and fitted distribution $P_\theta$ is measured by:
\begin{equation}
\begin{aligned}
d(P_\theta, Q) = \frac{1}{2} \{tr(&\Sigma_P^{-1}\Sigma_Q) - \log \det(\Sigma_P^{-1}\Sigma_Q) %\\
%&
+ (\mu_P - \mu_Q)' \Sigma_P^{-1}(\mu_P - \mu_Q) - n\}
\end{aligned}
\end{equation}
where $P_\theta$ and $Q$ define Gaussian distributions on $\Re^n$ with mean vectors $\mu_P$ and $\mu_Q$, respectively, and covariance matrices $\Sigma_P$ and $\Sigma_Q$, respectively. The KL-D in Table~\ref{table:sim} are on the collapsed space $\theta = \{ \beta, \sigma^2, \tau^2, \phi \}$. 
%We use the spatial linear regression model \eqref{eq: spatial_regression_model} and treat $\theta$ as the given parameters in the probability of $y$ to test the performance of the inference for $\theta$. 
We estimated the KL-D by the empirical estimator: 
\begin{equation}
E_{\theta \given y(S)}(d(P_\theta, Q)) \approx {1 \over L} \sum_{i = 1}^L d(P_{\theta_{(i)}}, Q)\;,
\end{equation}
where $\theta_{(i)}, i = 1, \ldots, L$ are $L$ samples from the posterior distribution of $\theta$. We also present the 95\% credible intervals for $d(P_{\theta}, Q)$ in Table~\ref{table:sim}.  The predicted outcome at any withheld location $s_0$ was estimated as
\begin{equation}
\hat{y}(s_0) = E[\tilde{y}(s_0) \given y(S)] \approx {1 \over L} \sum_{i = 1}^{L} \tilde{y}_{\theta_{(i)}}(s_0) \;,
\end{equation}
where $\tilde{y}_{\theta_{(i)}}(s_0) \sim p(y(s_0) \given y(S), \theta_{(i)})$ and $p(\cdot\given y(S), \theta_{(i)})$ is the likelihood for the respective model. These were used to calculate the RMSPE using the $200$ hold-out values.  We randomly picked 300 out of the 10000 samples from the post burn-in MCMC chains for calculating the KL-D and RMSPE. The $y(s_0)$ for full Gaussian process based and the latent NNGP model are sampled by function \textit{spPredict}. For the purpose of assessing the performance of recovering spatial latent process, we also report the Mean Squared Error (MSE) with respect to the true values of the spatial latent process (MSE($w$)) over the observed locations in the simulation. The KL-D, MSE($w$) and RMSPE metrics reveal that the NNGP provides a highly competitive alternative to the full Gaussian process based model. 

Table~\ref{table:sim} lists the parameter estimates and performance metrics for the candidate models. The posterior inference of the regression coefficients $\beta$ are close for all three models. While the posterior estimates of $\{\sigma^2. \tau^2, \phi \}$ are similar for full Gaussian process based model and latent NNGP model but, somewhat expectedly, different from the conjugate latent NNGP model. The 95\% confidence interval for $\sigma^2$ and $\tau^2$ are narrower since we fix the parameter $\phi, \delta^2$. The KL-Ds on the parameter space $\{w, \beta, \tau^2\}$ show that the conjugate latent NNGP provides reliable inference for the latent process and the regression coefficients. The same RMSPE across all three models also support that conjugate latent NNGP is comparable with full Gaussian process based model in prediction. The latent NNGP model is 200 times faster than the full Gaussian process based model, while the conjugate latent NNGP model use one tenth of the time required for the latent NNGP model to obtain similar inference on the regression coefficients and latent process. Notice that the time for the sampling of the 300 samples after fixing the parameter $\phi$ and $\delta^2$ in the conjugate latent NNGP model is less than one second. And the conjugate latent NNGP spare the effect of testing the tuning parameters in MCMC algorithm. Based on KL-D and RMSPE, the conjugate latent NNGP models emerge as highly competitive alternatives to latent NNGP models for prediction and inference on the latent process. 

\begin{figure}
	\centering
	\subfloat[True]{\includegraphics[width = 0.30\linewidth]{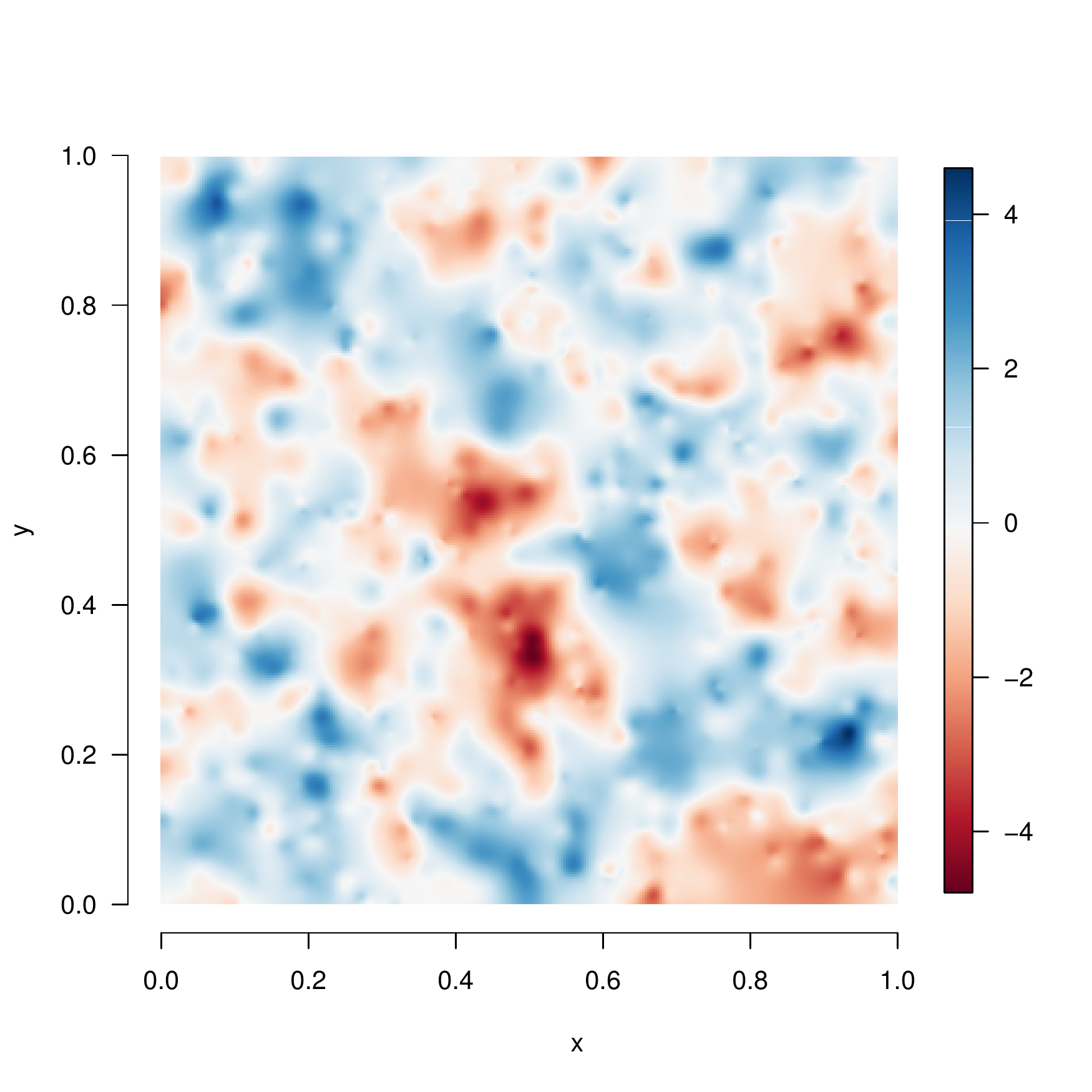}}
	\subfloat[fullGP]{\includegraphics[width = 0.30\linewidth]{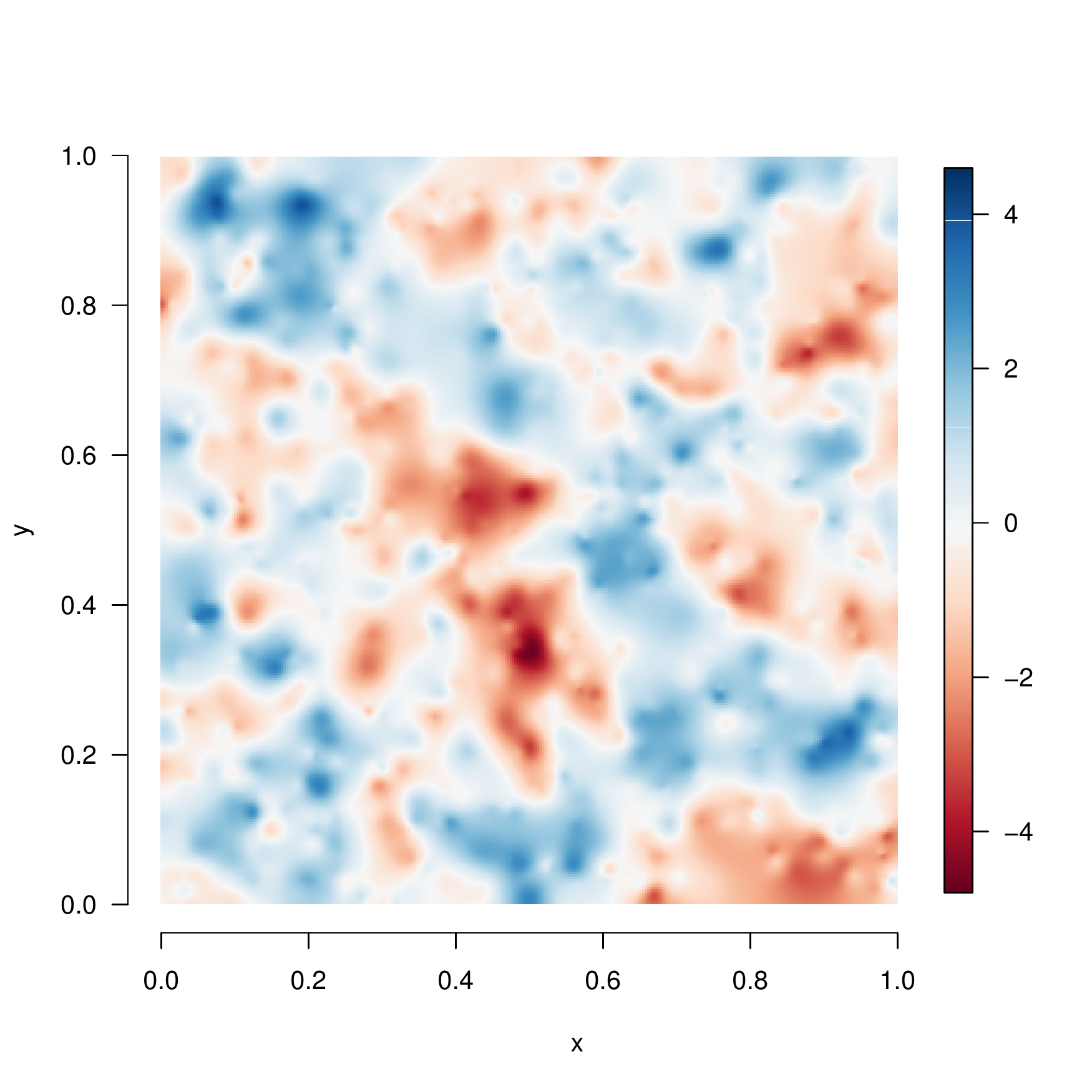}}
	\subfloat[Latent NNGP]{\includegraphics[width = 0.30\linewidth]{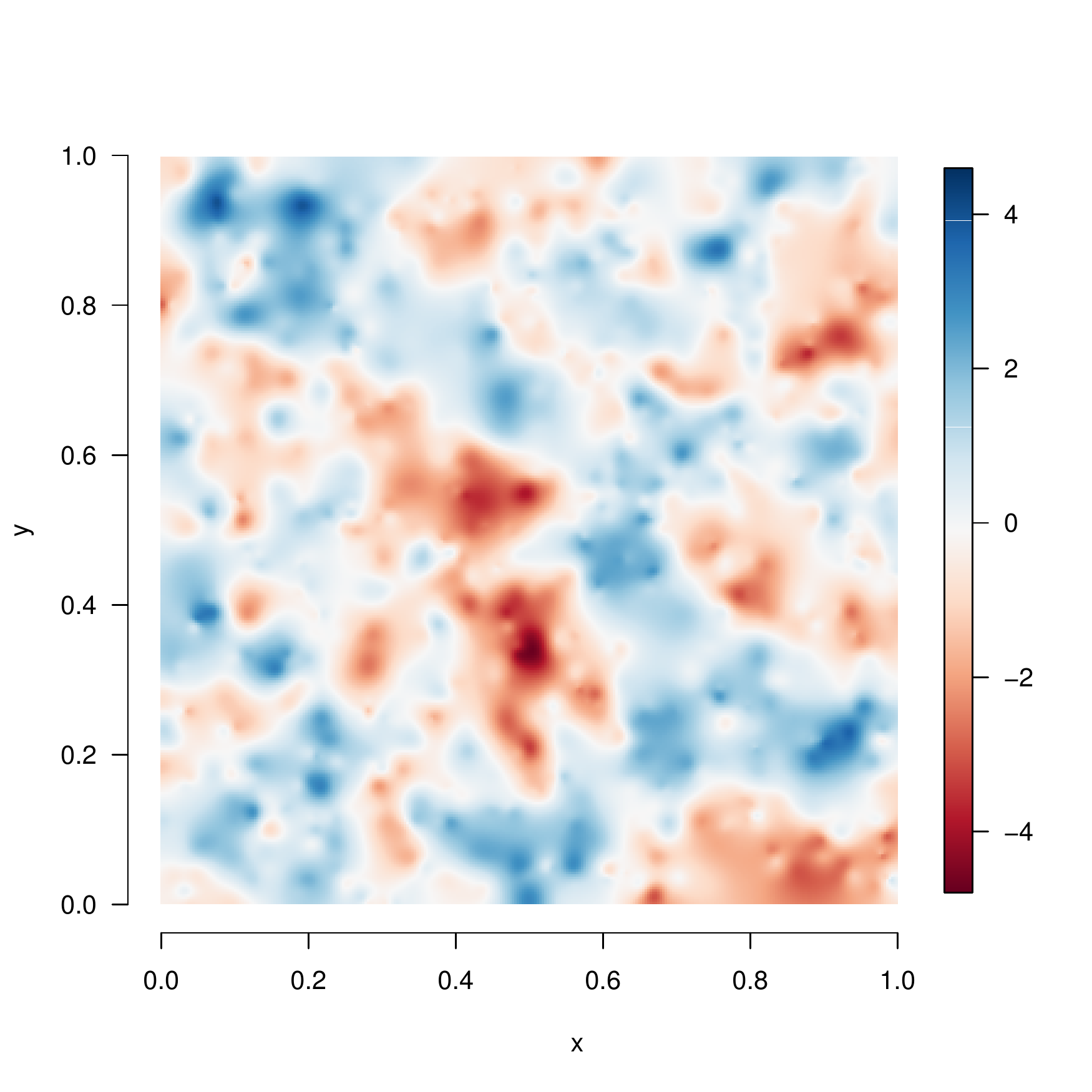}} \\
	\subfloat[Conjugate Latent NNGP]{\includegraphics[width = 0.30\linewidth]{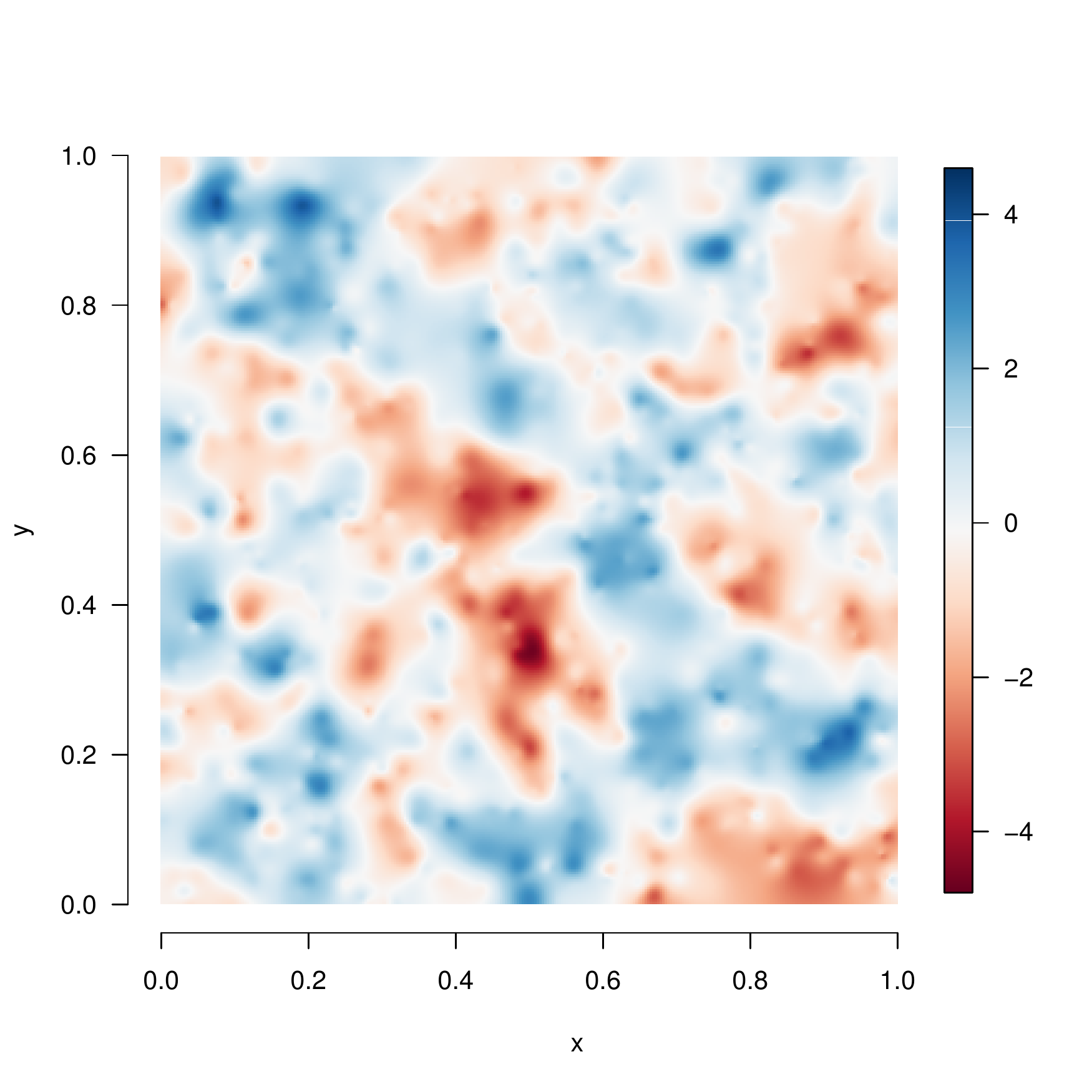}}
	\subfloat[CIs of w from fullGP\label{fig: CI_w_fullGP}]{\includegraphics[width = 0.30\linewidth]{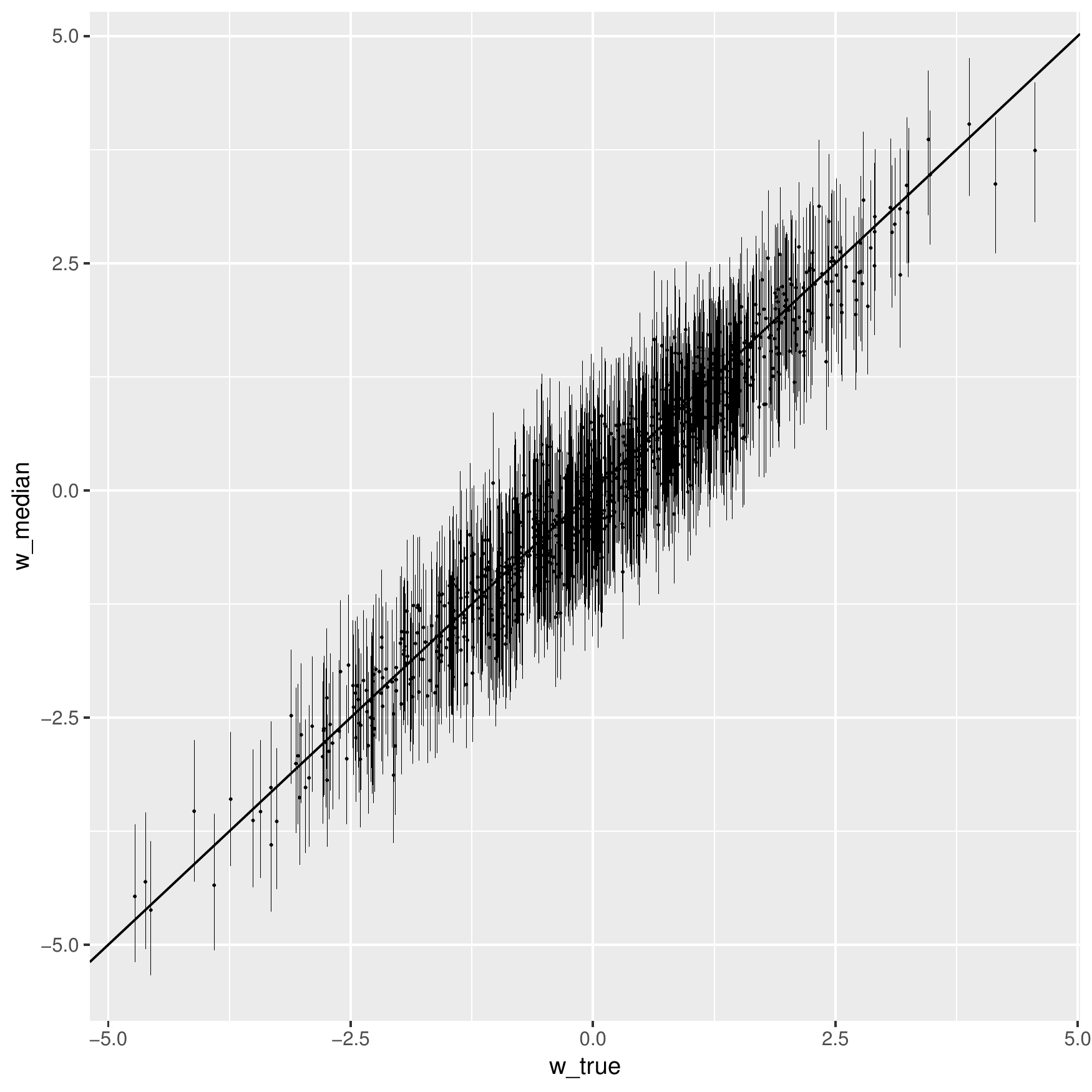}}
	\subfloat[CIs of w from Conjugate Latent NNGP \label{fig: CI_w_CLNNGP}]{\includegraphics[width = 0.30\linewidth]{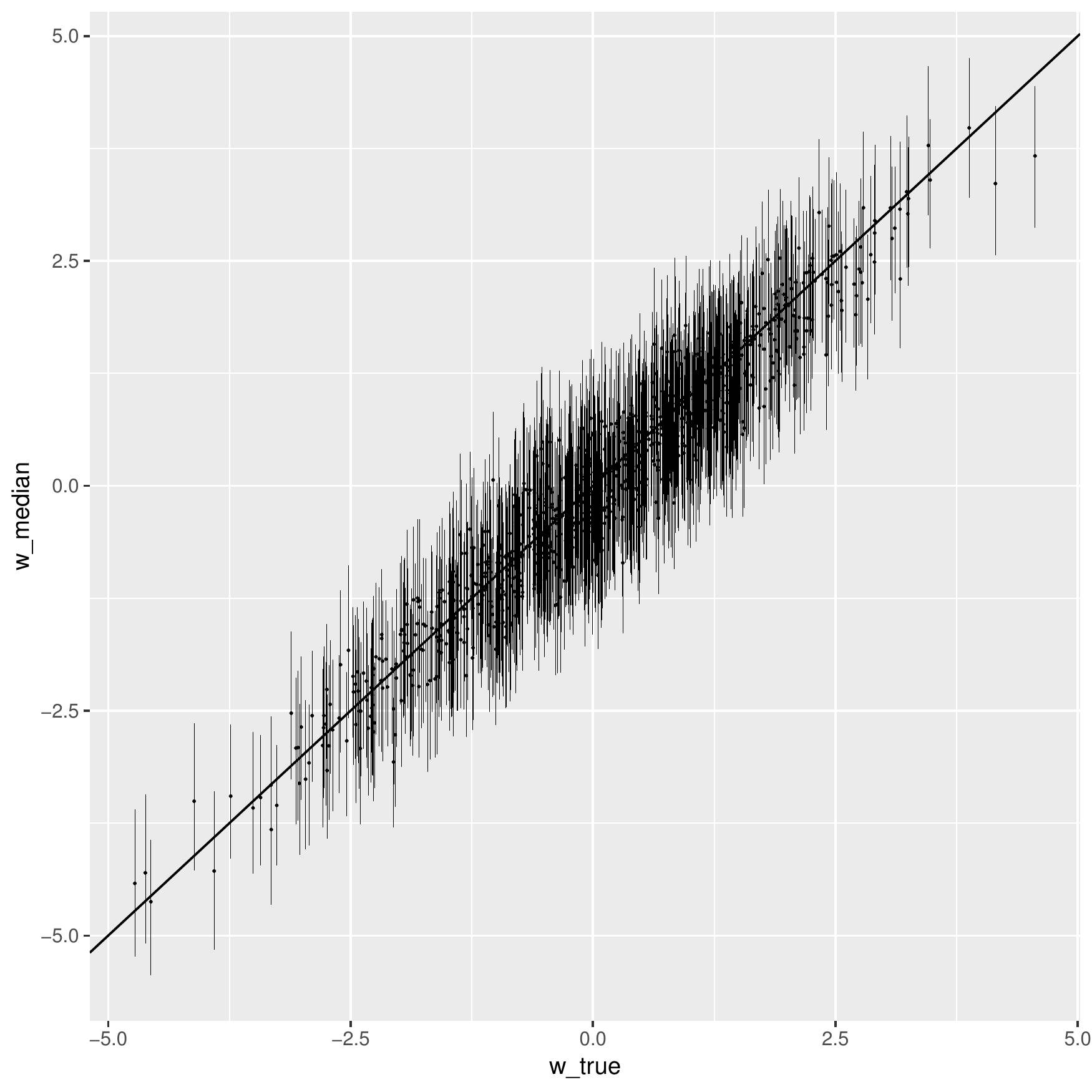}}
	\caption{Interpolated maps of (a) the true generated surface, the posterior means of the spatial latent process $w(s)$ for (b) the full Gaussian Process (Full GP), (c) the latent NNGP and (d) the conjugate latent NNGP. The 95\% confidence intervals for $w$ from (e) the full GP and (f) the conjugate latent NNGP. The models in (c), and (d) were all fit using $m=10$ nearest neighbors. \label{fig:w_com}} 
\end{figure}

%\subsection{Recovery of latent spatial process}\label{sec: recover_w}
Figure~\ref{fig:w_com} shows interpolated surfaces from the simulation example: \ref{fig:w_com}(a) shows an interpolated map of the ``true'' spatial latent process $w$, \ref{fig:w_com}(b)--(d) are maps of the posterior means of the latent process using a full GP model, a latent NNGP model and a conjugate latent NNGP model, respectively. Figure~\ref{fig:w_com}(e)--(f) present the 95\% confidence intervals for $w$ from a full GP model and a conjugate latent NNGP model. 
The recovered spatial residual surfaces are almost indistinguishable, and are comparable to the true interpolated surface of $w(s)$. Notice that the posterior mean of $w$ of the conjugate latent NNGP model can be theoretically calculated by the $\hat{\gamma}$ in \eqref{eq: joint_posterior_cojugate}. Thus the posterior samples of the latent process $w$ is only required for measuring uncertainty. Figure~\ref{fig: CI_w_CLNNGP} provides the 95\% confidence interval for all latent process $w$ from the conjugate latent NNGP model. There are 955 out of 1000 95\% confidence intervals successfully include the true value. This is comparable to the full Gaussian process based model (fig~\ref{fig: CI_w_fullGP}) which has 946 out of 1000 95\% confidence intervals covering the true value. 

\iffalse
quantile-quantile plots for the posterior distributions of $w(s_i)$'s from \textbf{Model1} and \textbf{Model2} against the latent NNGP model for three randomly selected locations.
\begin{figure}
	\centering
	\subfloat[$w_1$]{\includegraphics[width = 0.30\linewidth]{qq_pseodu_RE_w1}}
	\subfloat[$w_2$]{\includegraphics[width = 0.30\linewidth]{qq_pseodu_RE_w2}}
	\subfloat[$w_3$]{\includegraphics[width = 0.30\linewidth]{qq_pseodu_RE_w3}}\\
	\subfloat[$w_1$]{\includegraphics[width = 0.30\linewidth]{qq_conj_RE_w1}}
	\subfloat[$w_2$]{\includegraphics[width = 0.30\linewidth]{qq_conj_RE_w2}}
	\subfloat[$w_3$]{\includegraphics[width = 0.30\linewidth]{qq_conj_RE_w3}}
	\caption{Quantile-quantile plots of the posterior samples of $w(s)$ for 3 locations from \textbf{Model1} and \textbf{Model2} against those from the latent NNGP model\label{fig:NNGP_Conjugate_qq}}
\end{figure}	
The quantile-quantile plots indicate that \textbf{Model1} provides a slightly better approximation of the true posterior distribution than \textbf{Model2}, although \textbf{Model2} performs exact inference without resorting to MCMC and can rapidly deliver inference even for very large datasets. 
\fi

\section{Sea surface temperature analysis}\label{sec: real_data_analysis}
Global warming continues to be an ongoing concern among scientists. In order to develop conceptual and predictive global models, NASA monitors temperature and other atmospheric properties of the Earth regularly by two Moderate Resolution Imaging Spectroradiometer (MODIS) instruments in Aqua and Terra platforms. There is an extensive global satellite-based database processed and maintained by NASA. Details of the data can be found in \url{http://modis-atmos.gsfc.nasa.gov/index.html}. In particular, inferring on processes generating sea surface temperatures (SST) are of interest to atmospheric scientists studying exchange of heat, momentum, and water vapor between the atmosphere and ocean. %It is assumed that there exist unobserved factors such as ocean currents and tropical cyclones, which are potentially important explanatory variables to accurately predict spatial variations in SST.
Our aforementioned development will enable scientists to analyze large spatially-indexed datasets using a Bayesian geostatistical model easily implementable on modest computing platforms. 

\begin{figure}
	\centering
	\includegraphics[width = 0.4\linewidth]{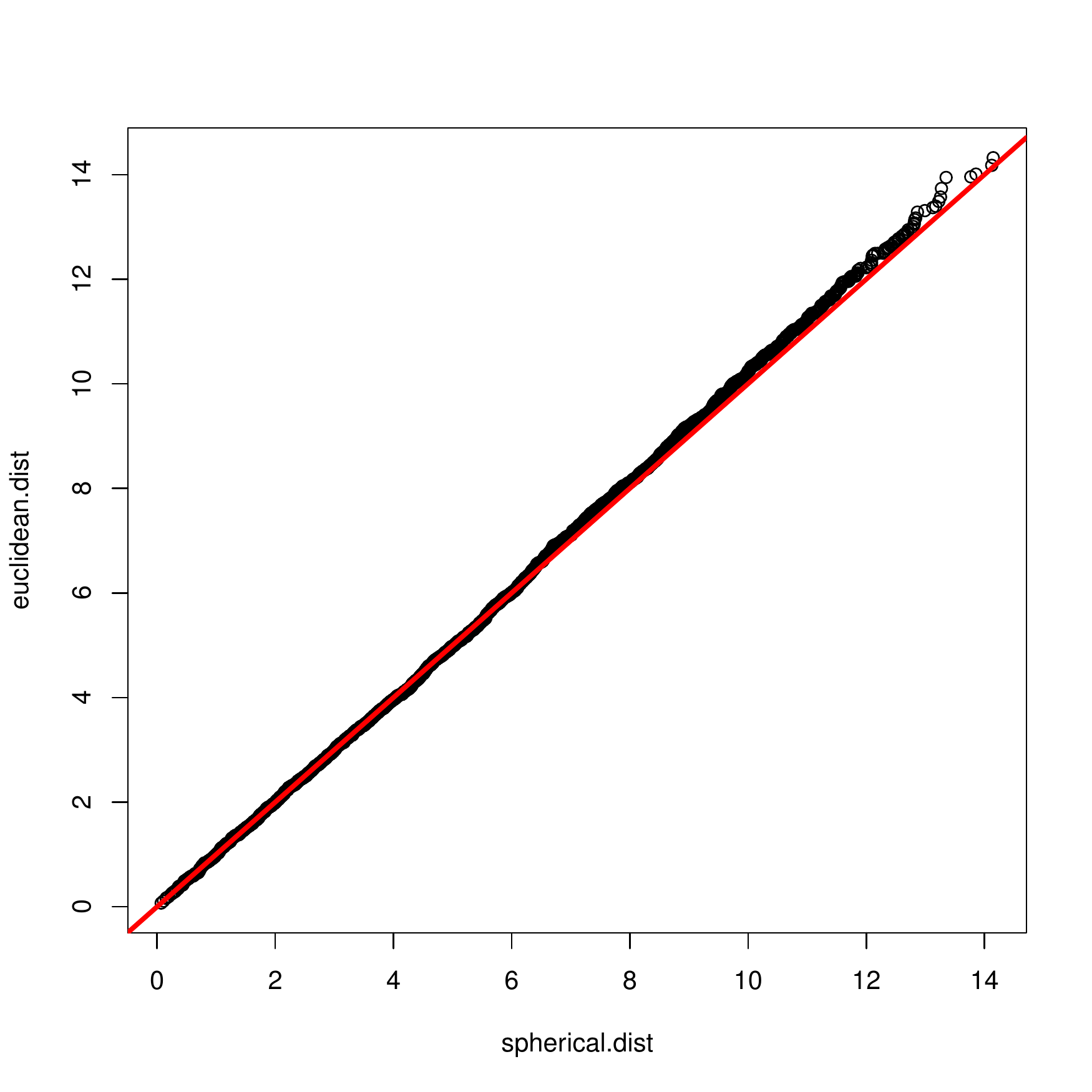}
	\caption{The Q-Q plot of the euclidean distance v.s. the spherical distance of 4000 pairs of observed locations over the study domain of the SST analysis. The red line is the 45 degree line}\label{fig: QQ_sinus_sphere}
\end{figure}

\begin{figure}
	\centering
	\subfloat[Observed SST over locations for training \label{fig: SST_train}]{\includegraphics[width = 0.45\linewidth]{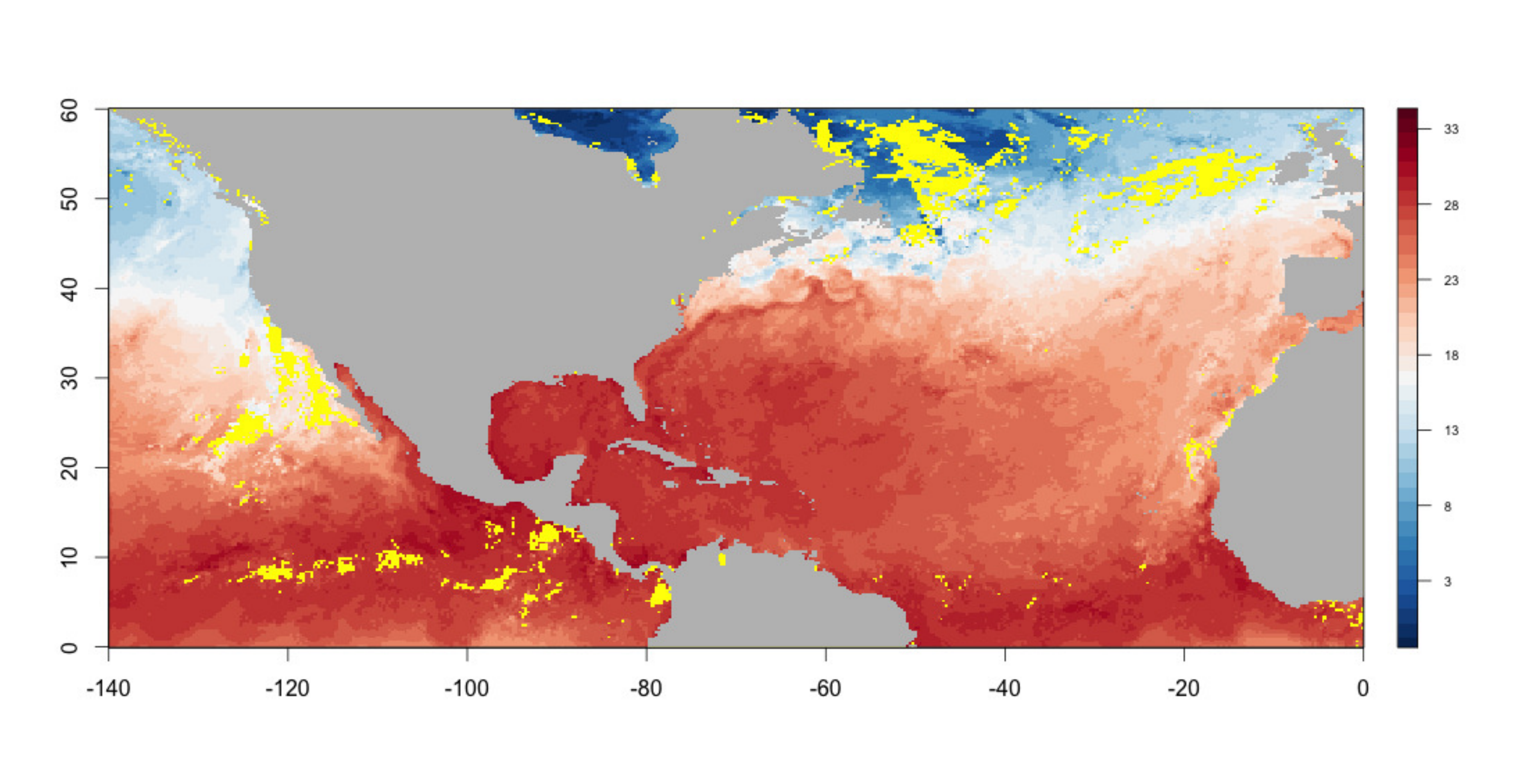}}
	\subfloat[Posterior mean of $w$ over locations for training by Conjugate latent NNGP\label{fig: post_mean_w_train}]{\includegraphics[width = 0.45\linewidth]{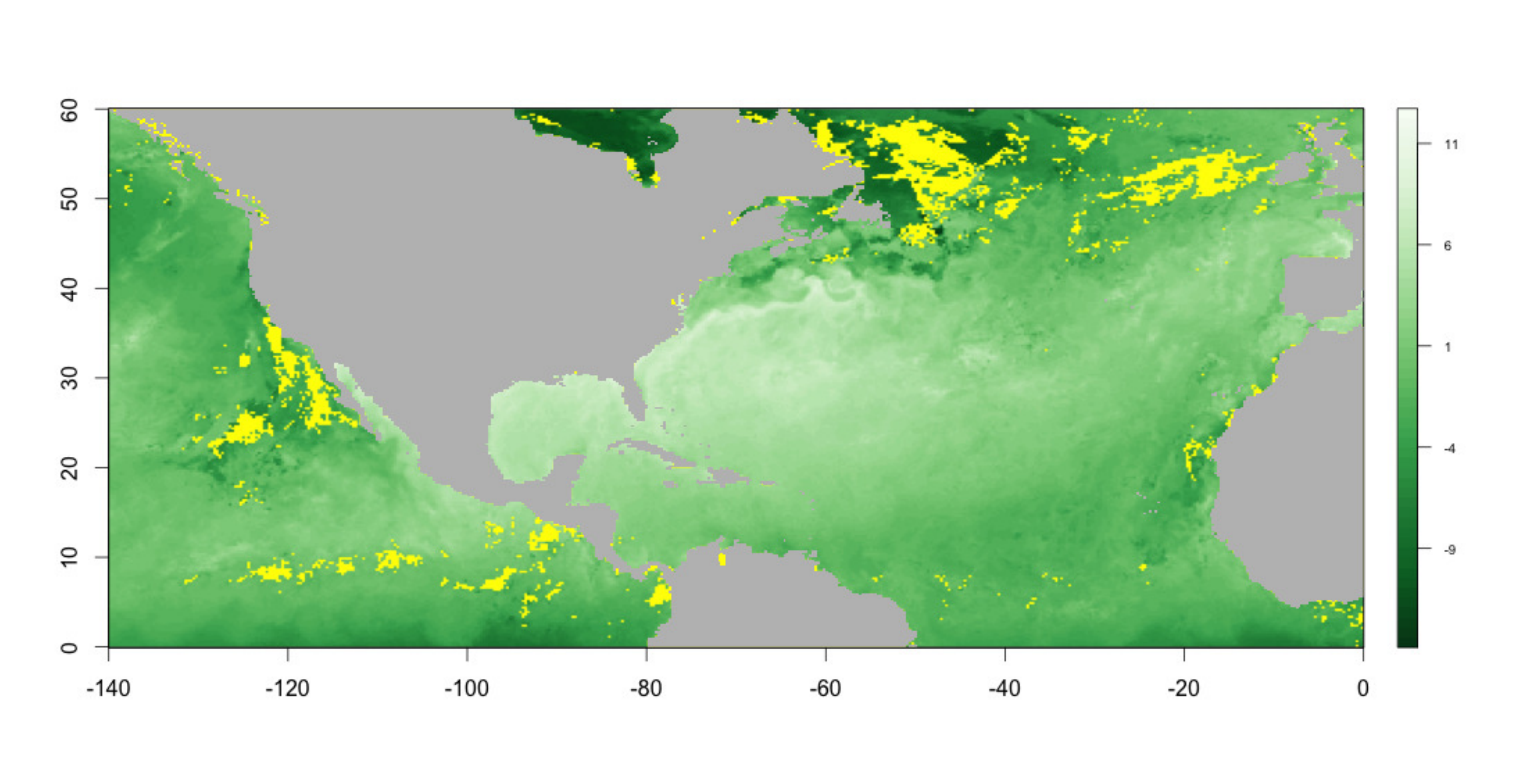}} \\
	\subfloat[Observed SST over locations for testing \label{fig: SST_test}]{\includegraphics[width = 0.45\linewidth]{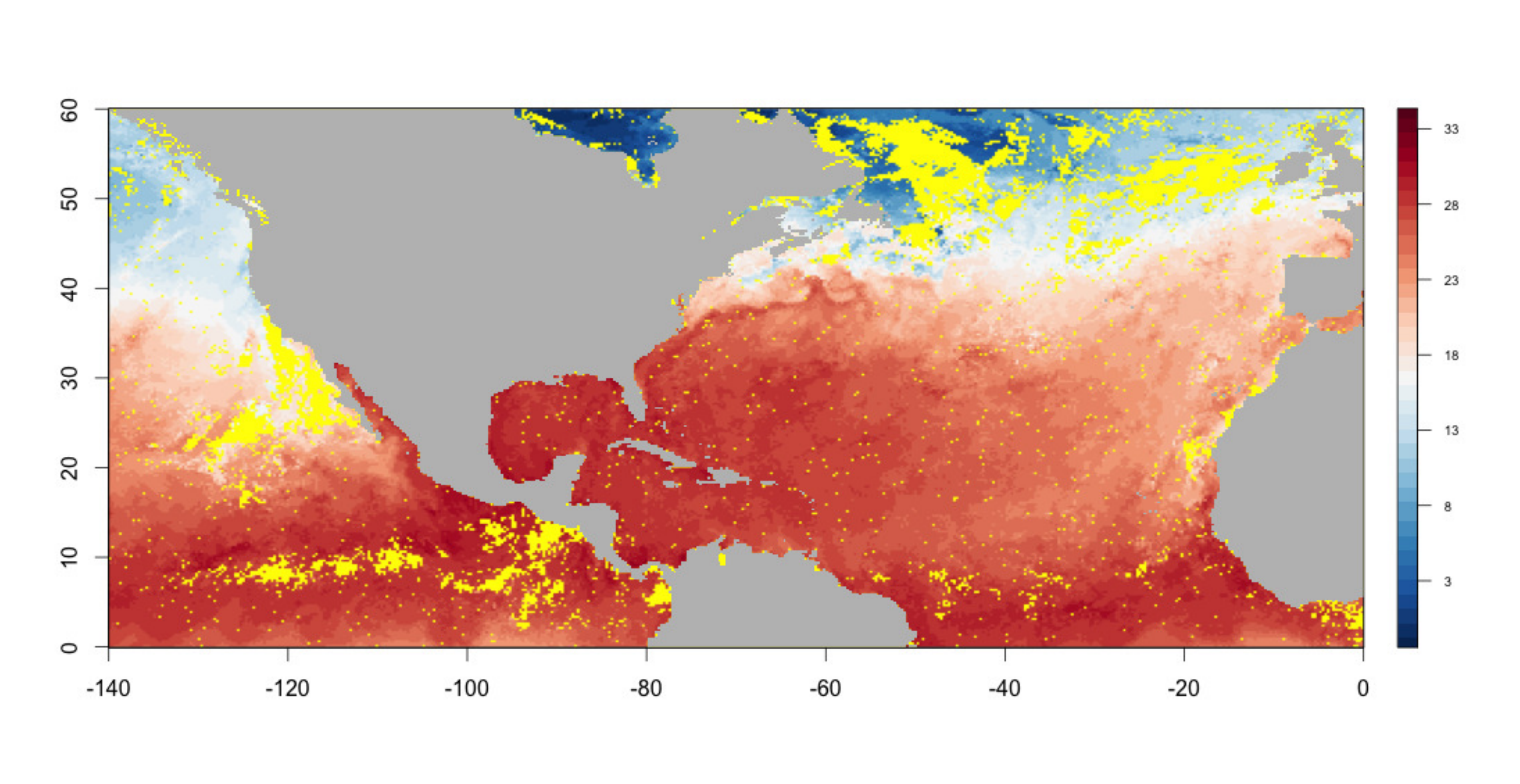}}
	\subfloat[Posterior mean of SST over locations for testing \label{fig: fitted_SST_test}]{\includegraphics[width = 0.45\linewidth]{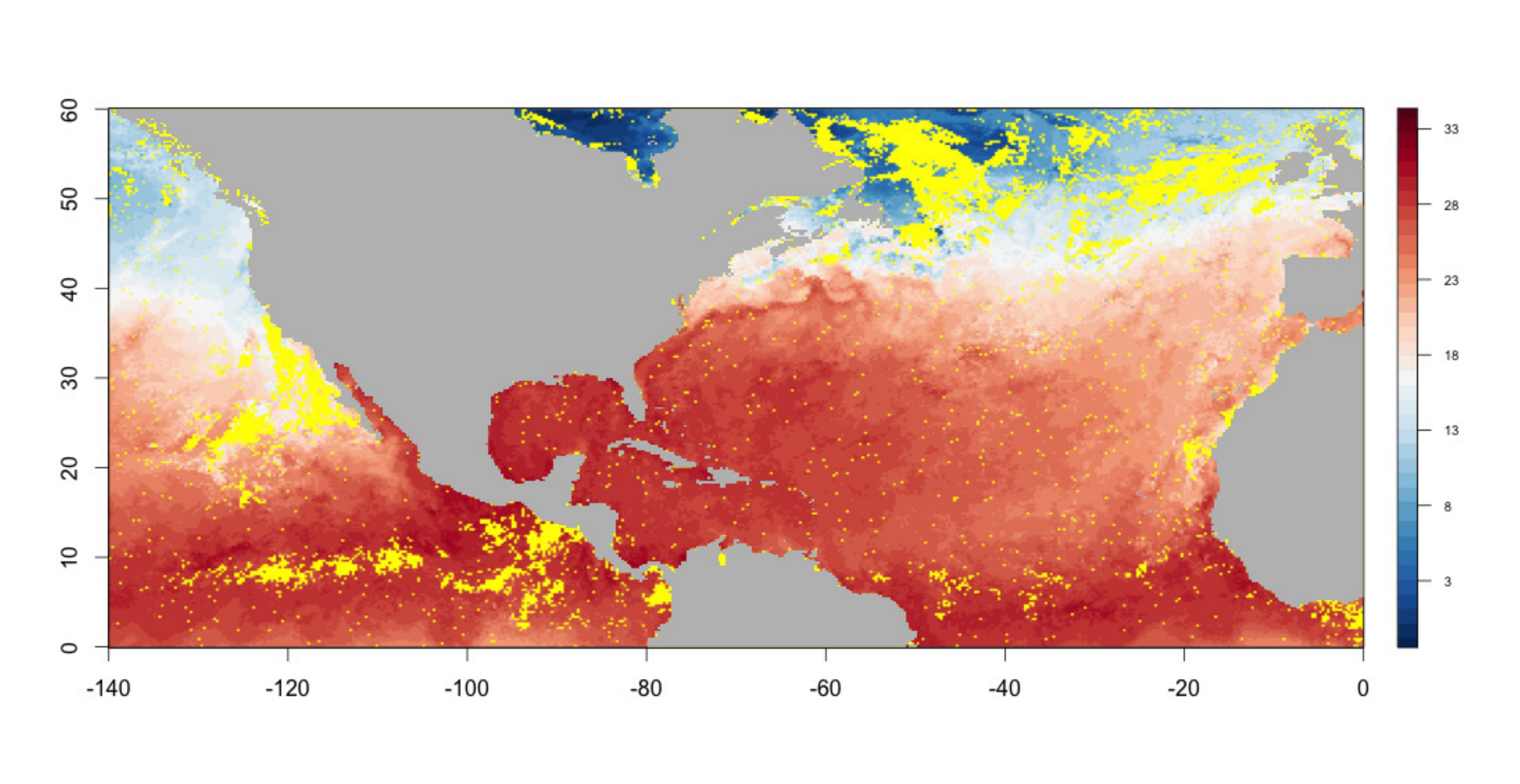}}
%	\subfloat[Ocean currents \label{fig: ocean_currents}]{\includegraphics[width = 0.45\linewidth]{ocean_current1}}
	\caption{Notes: (a) Observed SST over locations for training (b) Posterior mean of $w$ over locations for training by Conjugate latent NNGP (c) Posterior mean of SST over locations for testing (d) Posterior mean of SST over locations for testing The land is colored by gray, locations in the ocean without observations are colored by yellow. }
\end{figure}

Model-based inference is obtained rapidly using the conjugate latent NNGP model and, based on simulation studies, will be practically indistinguishable from MCMC-based output from more general NNGP specification. The dataset we analyze here consists of 2,827,252 spatially indexed observations of sea surface temperature (SST) collected between June 18-26, 2017, the data covers the ocean from longitude -140$^{\circ}$ to $^{\circ}$0 and from latitude 0$^{\circ}$ to 60$^{\circ}$. Among the 2,827,252 observations, $n = 2,544,527$ (90\%) were used for model fitting and the rest were withheld to assess predictive performance of the candidate models. Figure~\ref{fig: SST_train} depicts an interpolated map of the observed SST records over training locations. The temperatures are color-coded from shades of blue indicating lower temperatures, primarily seen in the higher latitudes, to shades of red indicating high temperatures. The missing data are colored by yellow and the gray part refers to land. To understand trends across the coordinates, we used sinusoidally projected coordinates (scaled to 1000km units) as explanatory variables. %\textcolor{blue}
{The sinusoidal projection is a popular equal-area projection [see, e.g., \cite{banerjee2005geodetic} or page~10 in \cite{banerjee2014hierarchical}]. We compare the Euclidean distances computed from a sinusoidal projection and the spherical or geodesic distance over the study domain by checking the two distances for 4000 pairs of locations randomly selected from the observed location set. The Q-Q plot (figure~\ref{fig: QQ_sinus_sphere}) shows that the Euclidean distance based on sinusoidal projects serves as a good measure of distance over the study domain.}
An exponential spatial covariance function %\textcolor{blue}
{with sinusoidally projected distance} was used for %\textcolor{blue}
{the model}. Further model specifications included non-informative flat priors for the intercept and regression coefficients, inverse-gamma priors for $\tau^2$ and $\sigma^2$ with shape parameter $2$ and scale parameter equaling the respective estimates from an empirical variogram. 
%The spatial decay $\phi$ was assigned a uniform prior $U(0.74, 74)$ suggesting a minimum spatial range of about $1\%$ of the maximum distance. 
%Since the memory of our workstation was insufficient to fit a Full GP model, 

We fit the conjugate Bayesian model with fixed $\phi$ and $\delta^2$
%the (non-conjugate) response NNGP model in (\ref{eq: nngp_response_posterior}) using \texttt{Stan} and the two conjugate Bayesian models, \textbf{Model1} and \textbf{Model2} described earlier, 
using the algorithm~1 in Section~3.3%\ref{sec: conjugate_nngp}
with $m=10$ nearest neighbors. We implement Algorithm~2 to choose the values of $\{ \phi, \delta^2 \}$ at $\phi = 7$, $\delta^2 = 0.001$. 
%For the response NNGP model, we generated 3 MCMC chains from \texttt{Stan} with 500 iterations each for the response NNGP model and took the first half as burn-in (default in \texttt{Stan}). The $\hat{R}$ statistics provided by \texttt{Stan} for all parameters were approximately equal to $1$ and subsequent inference was carried out using the $750$ posterior samples across the three chains. %We tested the inference of the response NNGP model by spNNGP. A MCMC chain with 25000 iterations from spNNGP provides a similar posterior inference. The posterior sample size for \textbf{Model1} and \textbf{Model2} were 750, which equals the number of samples got from response NNGP model.   
%The interpolation of the RMSPE is seen in Fig~ using the \texttt{spConjNNGP} function in \texttt{spNNGP} package in \texttt{R} indicated $\phi = 7$ and $\delta^2 = 0.001$ for \textbf{Model2}. %Although the choice of $\delta^2$ is on the edge of the interpolation, we choose $\delta^2$ down to 0.001 since a $\delta^2$ lower than $10^{-3}$ suggest that the nuggets is almost negligible to the partial sill. 
Figures~\ref{fig: post_mean_w_train} shows the posterior means for the latent process of the conjugate latent NNGP model. The temperatures are color-coded from light green indicating high temperatures to dark of green indicating low temperatures. The map of the latent process $w$ indicates lower temperature on the east coast and higher temperature on the west coast. At the same time, we observed high temperture at center of the map. These features coincide with the ocean current, suggesting that the ocean current plays an important role in the sea surface temperature.

%Very similar spatial patterns are evinced from both these models; both suggest high SST along the coast and far from the coast

Parameter estimates along with their estimated 95\% credible intervals and performance metrics for candidate models are shown in Table~\ref{table:real_data}. 
% \begin{table} 
% 	\centering
% 	\def\~{\hphantom{0}}
% 	\caption{Real data analysis summary table. Parameter Posterior summary mean (2.5, 97.5) percentiles} 
% 	\label{table:real_data}
% 	\rule{0.49\textwidth}{1pt}
% 	\begin{tabular*}{0.49\textwidth}
% 		{@{}c@{\extracolsep{\fill}}c@{\extracolsep{\fill}}c@{\extracolsep{\fill}}c@{\extracolsep{\fill}}@{}}
% 		\hline
% 		&Non-spatial & Conjugate latent NNGP \footnote{m = 10} \\
% 		\hline
% 		$\beta_0$ 	& 31.92(31.91, 31.92) & 31.43 (31.28, 31.59)\\ 
% 		$\beta_1$ 	& 0.12 (0.12, 0.12) &  0.07  (0.05,  0.09) \\ 
% 		$\beta_2$ 	&-3.07 (-3.07, -3.07) & -3.03 (-3.08, -2.99)\\ 
% 		$\sigma^2$ & -- & 3.95  (3.94,  3.95) \\
% 		$\phi$ & --  &  7.00 \\ 
% 		$\tau^2$ & 11.44 (11.43, 11.46) & 3.95$e^{-3}$ (3.94$e^{-3}$,  3.95$e^{-3}$)\\ 
% 		\hline
% 		RMSPE & 1.13 &0.31  \\
% 		\hline
% %		& \textbf{Model1} & \textbf{Model2} \\
% %		\hline
% %		$\beta_0$  &  29.98(29.11, 30.86) & 30.05(29.24, 30.86)\\ 
% %		$\beta_1$  & -0.95(-1.27, -0.64) & -0.96(-1.26, -0.69)\\ 
% %		$\beta_2$  &  -3.18(-3.38, -2.99) & -3.20(-3.38, -3.03)\\ 
% %		$\sigma^2$  &  1.74(1.60, 1.89)  & 1.61 (1.60, 1.62)\\
% %		$\phi$ &   13.25 (12.16, 14.41) & 14 \\ 
% %		$\tau^2$ & 	
% %		1.98$e^{-4}$ (1.31$e^{-4}$,  2.88$e^{-4}$) &1.61$e^{-3}$(1.60$e^{-3}$, 1.62$e^{-3}$) \\ 
% %		\hline
% %		RMSPE & 0.26& 0.26 \\
% %		time(h)  & 0.44 & 0.12\\
% 	\end{tabular*}
% 	\rule{0.49\textwidth}{1pt}
% \end{table}
\begin{table} 
	\centering
	\def\~{\hphantom{0}}
	\caption{Real data analysis summary table. Parameter Posterior summary mean (2.5, 97.5) percentiles} 
	\label{table:real_data}
	\begin{tabular*}{\textwidth}
		{@{}c@{\extracolsep{\fill}}c@{\extracolsep{\fill}}c@{\extracolsep{\fill}}c@{\extracolsep{\fill}}@{}}
		\hline
		&Non-spatial & Conjugate latent NNGP \footnote{m = 10} \\
		\hline
		$\beta_0$ 	& 31.92(31.91, 31.92) & 31.43 (31.28, 31.59)\\ 
		$\beta_1$ 	& 0.12 (0.12, 0.12) &  0.07  (0.05,  0.09) \\ 
		$\beta_2$ 	&-3.07 (-3.07, -3.07) & -3.03 (-3.08, -2.99)\\ 
		$\sigma^2$ & -- & 3.95  (3.94,  3.95) \\
		$\phi$ & --  &  7.00 \\ 
		$\tau^2$ & 11.44 (11.43, 11.46) & 3.95$e^{-3}$ (3.94$e^{-3}$,  3.95$e^{-3}$)\\ 
		\hline
		RMSPE & 3.39 &0.31  \\
		\hline
%		& \textbf{Model1} & \textbf{Model2} \\
%		\hline
%		$\beta_0$  &  29.98(29.11, 30.86) & 30.05(29.24, 30.86)\\ 
%		$\beta_1$  & -0.95(-1.27, -0.64) & -0.96(-1.26, -0.69)\\ 
%		$\beta_2$  &  -3.18(-3.38, -2.99) & -3.20(-3.38, -3.03)\\ 
%		$\sigma^2$  &  1.74(1.60, 1.89)  & 1.61 (1.60, 1.62)\\
%		$\phi$ &   13.25 (12.16, 14.41) & 14 \\ 
%		$\tau^2$ & 	
%		1.98$e^{-4}$ (1.31$e^{-4}$,  2.88$e^{-4}$) &1.61$e^{-3}$(1.60$e^{-3}$, 1.62$e^{-3}$) \\ 
%		\hline
%		RMSPE & 0.26& 0.26 \\
%		time(h)  & 0.44 & 0.12\\
		\hline
	\end{tabular*}
\end{table}
The RMSPE for a non-spatial linear regression model, conjugate latent NNGP model were 1.13, 0.31, respectively. Compared to the spatial models, the non-spatial models have substantially higher values of RMSPE, which suggest that coordinates alone does not adequately capture the spatial structure of SST. The fitted SST map over the withheld locations (Fig~\ref{fig: fitted_SST_test}) using conjugate latent NNGP model is almost indistinguishable from the real SST map (Fig~\ref{fig: SST_test}). All the inference from the conjugate latent NNGP model are based on 300 samples. The sampling process took 2367 seconds. In average, the posterior mean of the latent process $w$ can be obtained within 20 seconds.  
%The fitted SST map using \textbf{Model1} is also indistinguishable from Fig~\ref{fig: fitted_SST_model2} and is not shown. These results indicate strong consistency in practical inference between \textbf{Model1} and \textbf{Model2}. Yet the latter takes around $\sim$7 minutes on a single processor, while the former takes $\sim$26 minutes ($\sim 73\%$ savings in CPU time). 

\section{Conclusions and Future Work}\label{sec: conclusion}
This article has attempted to address some practical issues encountered by scientists and statisticians in the hierarchical modeling and analysis for very large geospatial datasets. Building upon some recent work on nearest-neighbor Gaussian processes for massive spatial data, we build conjugate Bayesian spatial regression models and propose strategies for rapidly deliverable inference on modest computing environments equipped with user-friendly and readily available software packages. In particular, we have demonstrated how judicious use of a conjugate latent NNGP model can be effective for estimation and uncertainty quantification of latent (underlying) spatial processes. This provides an easily implementable practical alternative to computationally onerous Bayesian computing approaches. All the computations done in the paper were implemented on a standard desktop using \texttt{R} and \texttt{Stan}. The article intends to contribute toward innovations in statistical practice rather than novel methodologies.

%\textcolor{blue}
{%The attempt of implementing CG is shown to be efficient. % including the simulation study (see section~\ref{sec: simulation_study}) and the real data analysis (see section~\ref{sec: real_data_analysis}).
The subsequent research of speeding up Algorithm~1 will include the following two aspects. Firstly, the speed of convergence of the regular CG algorithm to the solution of a symmetric positive definite linear system $Ax = b$ depends on the condition number of the matrix $A$. In practice, a \emph{preconditioned} CG is much more beneficial. Preconditioning of the CG method in Algorithm~1 is achieved by using a symmetric positive definite preconditioner matrix, say $M = LL^{\top}$, to solve $\tilde{A}\tilde{x} = \tilde{b}$, where $\tilde{A} = L^{-1}AL^{-\top}$ and $\tilde{b} = L^{-1}b$. The solution for $Ax=b$ is then obtained as $x = L^{-\top}\tilde{x}$. 
%\textcolor{red}{Though the conjugate gradient solver with a Jacobi preconditioner performances well in our studies, t}
The preconditioner should be chosen carefully. It should enjoy high memory efficiency and also ensure that $\kappa(\tilde{A})$ is close to 1, where $\kappa(\cdot)$ denotes the condition number of a matrix. Without these conditions, the benefits of preconditioning will not be evident and further investigations are needed to specify efficient preconditioners for modifying Algorithm~1. %Particularly, it's inverse should have sparse square root and can capture the essence of the inverse of $X_\ast^\top X_\ast$.%, which can be viewed as an approximation of $[\frac{1}{\delta^2}I_n + M^{-1}]^{-1}$.
The second aspect is parallel computing. The posterior samples generated by Algorithm~1 are independent, allowing the possibility of generating them simultaneously. One could explore the use of different parallel programming paradigms such as message parsing interfaces and GPUs to dramatically reduce the sampling times in Algorithm~1. }

It is important to recognize that the conjugate Bayesian models outlined here are not restricted to the NNGP. Any spatial covariance structure that leads to efficient computations can, in principle, be used. There are a number of recently proposed approaches that can be adopted here. These include, but are not limited to, multi-resolution approaches \citep[e.g.,][]{Nychka_Wikle_Royle_2002,nychka2015,katzfussmultires}, covariance tapering and its use in full-scale approximations \citep[e.g.,][]{fur06,sang12,katzfuss2013}, and stochastic partial differential equation approximations \citep[][]{lindgrenruelindstrom2011}, among several others \citep[see, e.g.,][and references therein]{banerjee2017high}. 

With regard to the NNGP specifically, our choice was partially dictated by its easy implementation in \texttt{R} using the \texttt{spNNGP} package and in \texttt{Stan} as described in \url{http://mc-stan.org/users/documentation/case-studies/nngp.html}. The NNGP is built upon a very effective likelihood approximation \citep{ve88,stein04}, which has also been explored recently by several authors in a variety of contexts \citep[][]{stroud17,guinness16}. \cite{guinness16} provides empirical evidence about Vecchia's approximation outperforming other alternate methods, but also points out some optimal methods for permuting the order of the spatial locations before constructing the model. His methods for choosing the order of the locations can certainly be executed prior to implementing the models proposed in this article. Finally, an even more recent article by \cite{katzfuss2017general} proposes further extensions of the Vecchia approximation, but its practicability for massive datasets on modest computing environments with easily available software packages is yet to be ascertained.

\section*{Acknowledgements}
The authors wish to thank Dr. Michael Betancourt, Dr. Bob Carpenter and Dr. Aki Vehtari of the \texttt{STAN Development Team} for useful guidance regarding the implementation of non-conjugate NNGP models in \texttt{Stan} for full Bayesian inference. The work of the first and third authors was supported, in part, by federal grants NSF/DMS 1513654, NSF/IIS 1562303 and NIH/NIEHS 1R01ES027027.  

\section*{Supplementary Material}
All computer programs implementing the examples in this article can be found in the public domain and downloaded from \url{https://github.com/LuZhangstat/ConjugateNNGP}.

\appendix

%\nocite{*}% Show all bib entries - both cited and uncited; comment this line to view only cited bib entries;
%\bibliography{wileyNJD-AMS}%
\bibliography{lubib}

% \section*{Author Biography}
% 
% \begin{biography}{\includegraphics[width=60pt,height=70pt,draft]{empty}}{\textbf{Author Name.} This is sample author biography text this is sample author biography text this is sample author biography text this is sample author biography text this is sample author biography text this is sample author biography text this is sample author biography text this is sample author biography text this is sample author biography text this is sample author biography text this is sample author biography text this is sample author biography text this is sample author biography text this is sample author biography text this is sample author biography text this is sample author biography text this is sample author biography text this is sample author biography text this is sample author biography text this is sample author biography text this is sample author biography text.}
% \end{biography}

\end{document}